\documentclass[aps,prd,twocolumn,superscriptaddress,nofootinbib]{revtex4-2}

\usepackage{graphicx}
\usepackage{amsmath,amssymb}
\usepackage{hyperref}
\usepackage{dcolumn}
\usepackage{bm}
\usepackage{booktabs}
\usepackage{multirow}
\usepackage{tcolorbox}
\usepackage{slashed}
\usepackage{geometry}
\usepackage{array}
\usepackage{color}
\usepackage{xcolor}
\usepackage{relsize}
\usepackage{lipsum}
\usepackage{longtable}
\usepackage{appendix}
\usepackage{subfigure}
\usepackage[sort&compress]{natbib}
\usepackage{placeins}
\begin{document}

\title{Global Electroweak Fit Constraints on the Two-Higgs-Doublet Model in Light of the CDF $W$-Boson Mass}

\author{Hindi Zouhair}
\email{hindizouhair@gmail.com}
\thanks{ corresponding author}
\affiliation{Laboratory of High Energy Physics:(LHEP-MS), Mohammed V University, Rabat, Morocco}


\date{\today}

\begin{abstract}
The recent measurement of the $W$ boson mass by the CDF II collaboration exhibits a significant tension with the Standard Model (SM) prediction and other experimental determinations. In this work, we investigate the implications of this result within the framework of the Two-Higgs-Doublet Model (2HDM), focusing on radiative corrections to electroweak precision observables parameterized in terms of the oblique parameters $\Delta S$, $\Delta T$, and $\Delta U$. Using global electroweak fits, we analyze how the inclusion of the CDF measurement modifies the preferred parameter space. We show that the observed shift in $m_W$ can be accommodated in the 2HDM through enhanced contributions to $\Delta T$, arising from mass splittings in the scalar sector. The resulting constraints on the scalar spectrum are presented and compared with those obtained using previous electroweak data. These results highlight the role of precision observables in probing extended Higgs sectors and provide updated bounds on viable 2HDM parameter space.
\end{abstract}



\maketitle

\section{Introduction}

Precision measurements have long served as powerful probes of the Standard Model (SM), offering sensitivity to virtual effects of particles beyond the direct reach of current colliders. Among these observables, the mass of the $W$ boson, $m_W$, plays a central role in constraining electroweak dynamics through radiative corrections. The recent high-precision determination by the CDF II collaboration, $m_W = 80.4335 \pm 0.0094~\mathrm{GeV}$~\cite{CDF:2022}, shows a notable tension with the Standard Model prediction as well as with previous global electroweak analyses~\cite{Baak:2012kk,Baak:2014ora,Haller:2018nnx,DeBlas:2019okz}. shows a notable tension with the SM prediction as well as with previous experimental measurements. This discrepancy motivates a careful reassessment of its implications within well-motivated extensions of the SM.

Within the SM, $m_W$ is not a free parameter~\cite{Awramik:2003rn}, but is determined by a set of precisely measured inputs, including the top-quark mass, the Higgs boson mass, and the running of the electromagnetic coupling. Consequently, any significant deviation in $m_W$ can be interpreted as a signal of additional loop contributions from new particles or interactions. Such effects can be systematically analyzed within the framework of electroweak precision observables and global fit techniques.

In this work, we investigate the impact of the CDF measurement within the framework of the Two-Higgs-Doublet Model (2HDM), a minimal extension of the scalar sector of the SM. The 2HDM introduces an additional $SU(2)_L$ scalar doublet, leading to a richer Higgs spectrum and new contributions to electroweak observables at the loop level. These effects can be efficiently captured through the oblique parameters $S$, $T$, and $U$, which encode corrections to the gauge boson self-energies.

We compute the contributions of the extended scalar sector to the electroweak precision observables and analyze their impact on the global fit. Particular emphasis is placed on the parameter $\Delta T$, which is sensitive to custodial symmetry breaking and depends strongly on the mass splittings among the scalar states. By comparing the fit results obtained using the CDF measurement with those based on previous datasets, we identify the regions of parameter space in which the 2HDM can accommodate the observed shift in $m_W$.

The paper is organized as follows. In Sec.~II, we review the scalar sector of the Two-Higgs-Doublet Model. In Sec.~III, we summarize the oblique parameter formalism. In Sec.~IV, we describe the global electroweak fit. In Sec.~V, we present our results. Finally, we conclude in Sec.~VI.

\section{The Two-Higgs-Doublet Model}
The Two-Higgs-Doublet Model (2HDM) is a minimal extension of the Standard Model (SM) scalar sector obtained by introducing a second $SU(2)_L$ scalar doublet. In its CP-conserving realization with a softly broken $\mathbb{Z}_2$ symmetry, the most general renormalizable scalar potential can be written as~\cite{Branco:2011iw,Gunion:1989we,Haber:2010bw}
\begin{widetext}
	\begin{equation}
	\begin{aligned}
	V(\Phi_1,\Phi_2) &= m_{11}^2\,\Phi_1^\dagger \Phi_1 + m_{22}^2\,\Phi_2^\dagger \Phi_2
	- m_{12}^2 \left(\Phi_1^\dagger \Phi_2 + \Phi_2^\dagger \Phi_1 \right) \\
	&\quad + \frac{\lambda_1}{2}(\Phi_1^\dagger \Phi_1)^2
	+ \frac{\lambda_2}{2}(\Phi_2^\dagger \Phi_2)^2
	+ \lambda_3(\Phi_1^\dagger \Phi_1)(\Phi_2^\dagger \Phi_2) \\
	&\quad + \lambda_4(\Phi_1^\dagger \Phi_2)(\Phi_2^\dagger \Phi_1)
	+ \frac{\lambda_5}{2}\left[(\Phi_1^\dagger \Phi_2)^2 + (\Phi_2^\dagger \Phi_1)^2\right].
	\end{aligned}
	\label{eq:2hdm_potential}
	\end{equation}
\end{widetext}
After electroweak symmetry breaking, the two scalar doublets are expanded as
\begin{widetext}
	\begin{equation}
	\Phi_1 =
	\begin{pmatrix}
	\phi_1^+ \\
	\dfrac{1}{\sqrt{2}}(v_1 + \rho_1 + i\eta_1)
	\end{pmatrix},
	\qquad
	\Phi_2 =
	\begin{pmatrix}
	\phi_2^+ \\
	\dfrac{1}{\sqrt{2}}(v_2 + \rho_2 + i\eta_2)
	\end{pmatrix},
	\label{eq:doublets}
	\end{equation}
\end{widetext}
where $v_1$ and $v_2$ are the vacuum expectation values (VEVs), satisfying
\begin{equation}
v_1^2 + v_2^2 = v^2, \qquad v \simeq 246~\mathrm{GeV},
\end{equation}
and
\begin{equation}
\tan\beta = \frac{v_2}{v_1}.
\end{equation}
The physical scalar spectrum consists of two CP-even neutral Higgs bosons, $h$ and $H$, one CP-odd neutral scalar, $A$, and a charged Higgs pair, $H^\pm$. At tree level, the masses of the pseudoscalar and charged Higgs states are given by~\cite{Gunion:1989we,Eriksson:2009ws}
\begin{align}
m_A^2 &= \frac{m_{12}^2}{\sin\beta \cos\beta} - \lambda_5 v^2, \\
m_{H^\pm}^2 &= \frac{m_{12}^2}{\sin\beta \cos\beta} - \frac{\lambda_4 + \lambda_5}{2} v^2.
\end{align}
The CP-even mass eigenstates are obtained by diagonalizing the neutral scalar mass matrix with mixing angle $\alpha$, leading to the masses $m_h$ and $m_H$. In the phenomenologically relevant alignment limit,
\begin{equation}
\cos(\beta-\alpha) \to 0,
\end{equation}
the lighter CP-even state $h$ acquires SM-like couplings, consistent with current Higgs signal-strength measurements ~\cite{ATLAS:2012yve,CMS:2012qbp}.
This parametrization is standard in numerical studies and is implemented in public tools such as \texttt{2HDMC}~\cite{Eriksson:2009ws}. The contributions of the extended scalar sector to electroweak precision observables are governed primarily by the mass splittings among $H$, $A$, and $H^\pm$, which induce corrections to the oblique parameters, particularly $\Delta T$.
\vspace{-15pt}
\subsection{Oblique Parameters in the 2HDM}
Electroweak precision observables provide a sensitive indirect probe of physics beyond the SM. When new physics affects primarily the gauge-boson self-energies, its leading impact can be conveniently encoded in the oblique parameters $S$, $T$, and $U$~\cite{Peskin:1990zt,Peskin:1991sw,Altarelli:1990zd,Altarelli:1991fk}.. In this framework, the parameter $T$ is especially important, as it directly probes custodial-symmetry breaking and is therefore highly sensitive to mass splittings in extended scalar sectors such as the 2HDM.
The oblique parameters are defined in terms of the transverse vacuum-polarization amplitudes $\Pi_{VV'}(q^2)$. For example, the parameter $S$ can be written as
\vspace{25pt}
\begin{equation}
S = \frac{4 s_W^2 c_W^2}{\alpha}
\left[
\Pi_{ZZ}'(0)
- \frac{c_W^2-s_W^2}{s_W c_W}\Pi_{Z\gamma}'(0)
- \Pi_{\gamma\gamma}'(0)
\right],
\end{equation}
where
\begin{equation}
\Pi'_{VV'}(0) \equiv \left.\frac{\partial \Pi_{VV'}(q^2)}{\partial q^2}\right|_{q^2=0}.
\end{equation}
Analogous definitions apply to $T$ and $U$. In the 2HDM, the additional scalar states contribute to the electroweak gauge-boson self-energies at one-loop order. These corrections depend on the scalar masses and on the mixing angle $\beta-\alpha$, and can be expressed in terms of standard Passarino--Veltman two-point functions. In practice, the oblique corrections are most conveniently discussed in terms of the shifts
\vspace{-10pt}
\begin{equation}
\Delta X = X_{\rm 2HDM} - X_{\rm SM}, \qquad X \in \{S,T,U\},
\end{equation}
with the SM evaluated at the same reference Higgs-boson mass and electroweak input parameters.
\vspace{10pt}
\subsection{Analytic Expressions for \texorpdfstring{$\Delta S$, $\Delta T$, and $\Delta U$}{Delta S, Delta T, and Delta U}}
For completeness, we summarize the one-loop 2HDM contributions to the oblique parameters following standard electroweak precision analyses of extended scalar sectors~\cite{Grimus:2008nb}, using the notation of Ref.~\cite{Lu:2022}. These expressions are implemented numerically in our analysis.
\begin{widetext}
	\begin{equation}
	\begin{split}
	\Delta S = &\ \frac{\cos^2(\beta-\alpha)}{\pi m_Z^2}
	\Big[
	B_{22}(m_Z^2; m_H^2, m_A^2)
	- B_{22}(m_Z^2; m_{H^\pm}^2, m_{H^\pm}^2) \\
	&\quad + B_{22}(m_Z^2; m_h^2, m_A^2)
	- B_{22}(m_Z^2; m_H^2, m_A^2)
	+ B_{22}(m_Z^2; m_Z^2, m_H^2)
	- B_{22}(m_Z^2; m_Z^2, m_h^2) \\
	&\quad - m_Z^2 B_0(m_Z^2; m_Z^2, m_H^2)
	+ m_Z^2 B_0(m_Z^2; m_Z^2, m_h^2)
	\Big],
	\end{split}
	\end{equation}
	\begin{equation}
	\begin{split}
	\Delta T = &\ \frac{\cos^2(\beta-\alpha)}{16\pi m_W^2 s_W^2}
	\Big[
	F(m_{H^\pm}^2,m_A^2)
	+ F(m_{H^\pm}^2,m_H^2)
	- F(m_A^2,m_H^2) \\
	&\quad + F(m_{H^\pm}^2,m_h^2)
	- F(m_A^2,m_h^2)
	+ F(m_A^2,m_H^2)
	+ F(m_W^2,m_H^2)
	- F(m_W^2,m_h^2) \\
	&\quad + F(m_Z^2,m_h^2)
	- F(m_Z^2,m_H^2)
	+ 4m_Z^2 \bar{B}_0(m_Z^2; m_H^2, m_h^2)
	- 4m_W^2 \bar{B}_0(m_W^2; m_H^2, m_h^2)
	\Big],
	\end{split}
	\end{equation}
	\begin{equation}
	\begin{split}
	\Delta U = &\ -\Delta S
	+ \frac{\cos^2(\beta-\alpha)}{\pi m_W^2}
	\Big[
	B_{22}(m_W^2; m_A^2, m_{H^\pm}^2)
	- 2B_{22}(m_W^2; m_{H^\pm}^2, m_{H^\pm}^2) \\
	&\quad + B_{22}(m_W^2; m_H^2, m_{H^\pm}^2)
	+ B_{22}(m_W^2; m_h^2, m_{H^\pm}^2)
	- B_{22}(m_W^2; m_H^2, m_H^2)
	- B_{22}(m_W^2; m_h^2, m_h^2) \\
	&\quad - m_W^2 B_0(m_W^2; m_W^2, m_H^2)
	+ m_W^2 B_0(m_W^2; m_W^2, m_h^2)
	\Big].
	\end{split}
	\end{equation}
\end{widetext}

The numerical analysis presented below is based on these one-loop contributions, with particular emphasis on the dependence of $\Delta T$ on the scalar mass splittings. Since $\Delta T$ is especially sensitive to custodial-symmetry breaking~\cite{Veltman:1977kh}, it provides the dominant handle for assessing whether the 2HDM can accommodate the shift in the $W$-boson mass preferred by the CDF result.

\section{Electroweak Fit}

The global analysis of electroweak (EW) observables provides a precise framework for testing the internal consistency of the Standard Model (SM) and for quantifying possible deviations induced by new physics. In particular, the strong correlations among precision observables imply that changes in one input quantity, such as the $W$-boson mass, can significantly affect the preferred values of other parameters in a global fit. Following the methodology of precision-fit studies~\cite{Baak:2012kk,Baak:2014ora,Flacher:2008zq,Baak:2011ze,Haller:2018nnx,DeBlas:2019okz,deBlas:2021wap}, we perform a global electroweak fit incorporating precision measurements from LEP, SLD, Tevatron, and PDG compilations~\cite{Schael2006,Zyla:2020zbs,Aaltonen:2018dxj,CDF:2022}. 

\begin{figure*}[!t]
	\centering
	\includegraphics[width=\textwidth]{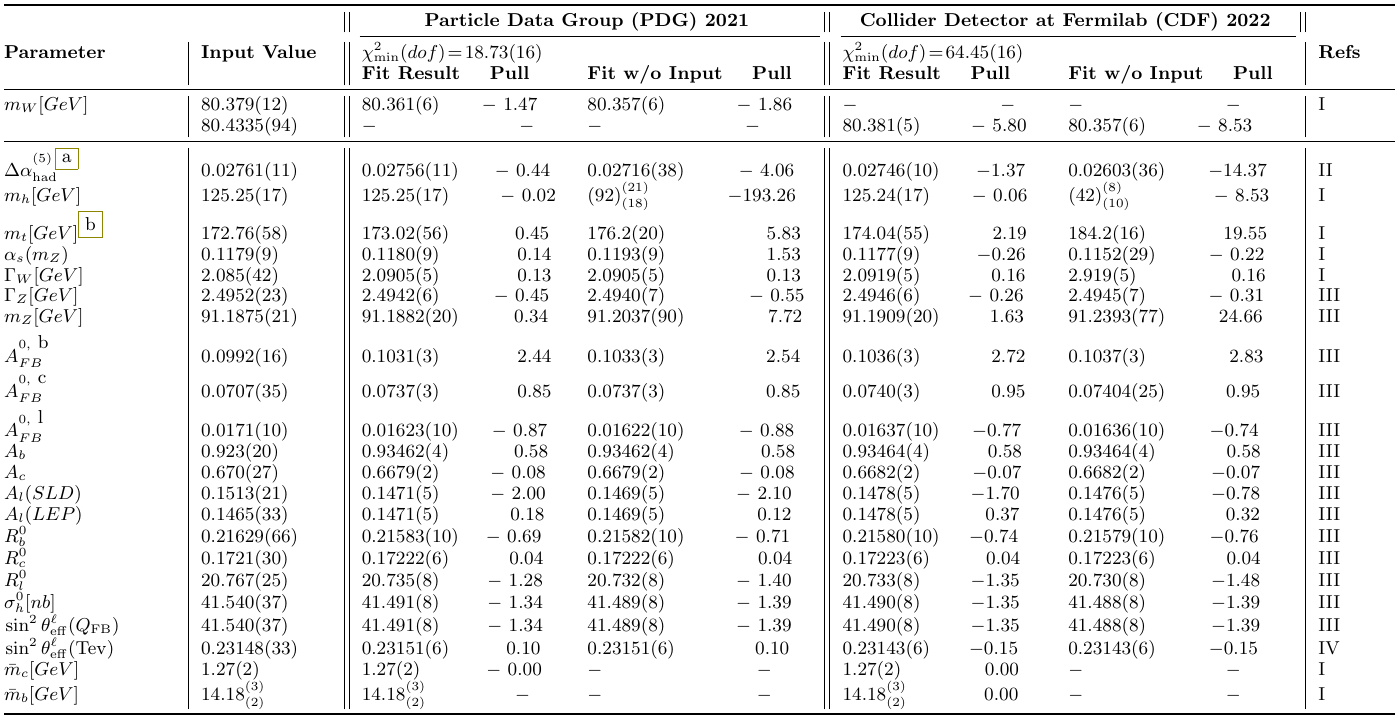}
	\caption{
		Input parameters and best-fit values from the global electroweak fit, with the Fermi constant 
		\(G_F = 1.1663787(6) \times 10^{-5} \, \mathrm{GeV}^{-2}\)~\cite{Zyla:2020zbs} fixed in the analysis. 
		Correlations among \((m_Z, \Gamma_Z, \sigma_h^0, R_\ell^0, A_\ell, A_{\mathrm{FB}}^\ell)\) and 
		\((A_{\mathrm{FB}}^c, A_{\mathrm{FB}}^b, A_c, A_b, R_c, R_b)\) are included as in Ref.~\cite{Schael2006}. 
		The “Pull” is defined as \((O_{\mathrm{fit}} - O_{\mathrm{measure}})/\sigma_{\mathrm{measure}}\), 
		where \(\sigma_{\mathrm{measure}}\) is the experimental uncertainty. 
		This table is adapted from Ref.~\cite{Lu:2022}. 
		$^{\fcolorbox{olive}{white}{a}}$ Computed with \(\alpha_s\) evaluated at the \(Z\) boson mass scale. 
		$^{\fcolorbox{olive}{white}{b}}$ A theoretical uncertainty of \(0.5\,\mathrm{GeV}\) has been included. 
		Roman source labels indicate the origin of the experimental inputs: 
		\(\mathrm{I}\)~\cite{Zyla:2020zbs}; 
		\(\mathrm{II}\)~\cite{Crivellin2020,Davier2020,Keshavarzi2020}; 
		\(\mathrm{III}\)~\cite{Schael2006}; 
		\(\mathrm{IV}\)~\cite{Aaltonen:2018dxj}.
	}
	\label{fig:EWfit}
\end{figure*}

\begin{figure*}[!t]
	\centering
	\includegraphics[scale=0.7]{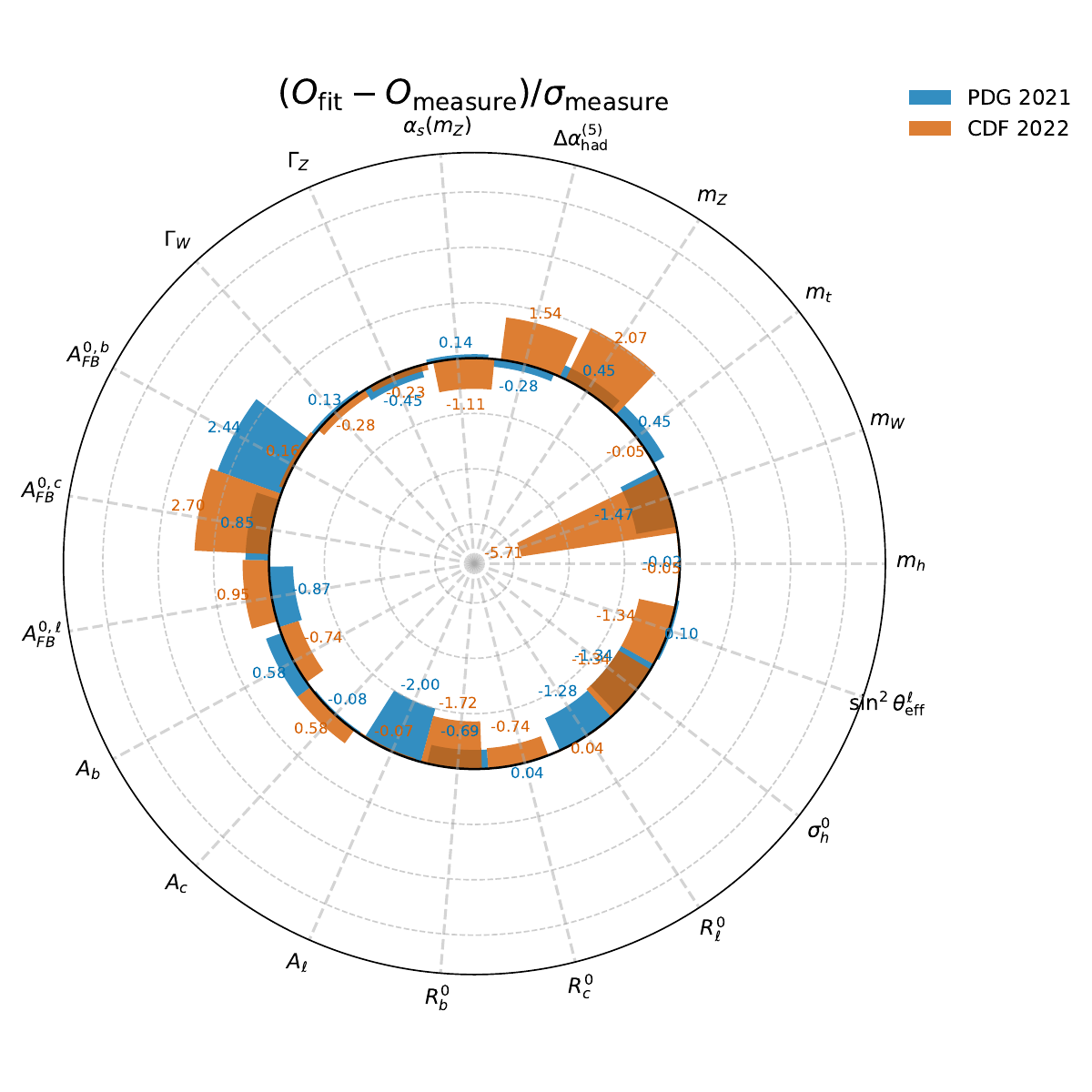}
	\caption{Comparison of pull values, $(O_{\mathrm{fit}}-O_{\mathrm{meas}})/\sigma_{\mathrm{meas}}$, for the main electroweak precision observables in global fits using the PDG 2021 (blue) and CDF 2022 (orange) values of the $W$-boson mass. The figure shows that the CDF result induces correlated shifts in multiple observables, rather than a single isolated deviation. This pattern is consistent with the need for positive oblique corrections, especially in $\Delta T$, as expected from custodial-symmetry-breaking effects in extended scalar sectors. The numerical pulls correspond to those shown in Fig.~\ref{fig:EWfit}.}
	\label{comparison_pdg_cdf}
\end{figure*}

To assess the impact of the CDF II result, we compare two fit setups using different reference values for the $W$-boson mass:
\begin{enumerate}
	\item the PDG 2021 average, $m_W = 80.379 \pm 0.012~\mathrm{GeV}$,
	\item the CDF II measurement, $m_W = 80.4335 \pm 0.0094~\mathrm{GeV}$~\cite{CDF:2022}.
\end{enumerate}
The resulting shifts in the fitted observables are illustrated in Fig.~\ref{fig:EWfit}. To visualize more clearly how the inclusion of the CDF $W$-boson mass measurement reshapes the global electroweak fit, we show in Fig.~\ref{comparison_pdg_cdf} the pull values of the main precision observables for the PDG 2021 and CDF 2022 input choices. This representation makes it possible to identify at a glance which observables are most strongly affected by the updated $m_W$ value and how the resulting tension propagates across the electroweak sector. In particular, the pattern of shifts indicates that the CDF anomaly cannot be regarded as an isolated deviation in $m_W$ alone, but instead induces correlated distortions in other precision observables such as $m_t$, $m_Z$, and $\Delta\alpha_{\mathrm{had}}^{(5)}(M_Z)$. This behavior is naturally suggestive of oblique corrections and, within extended scalar sectors such as the 2HDM, points especially toward a positive contribution to $\Delta T$.

According to the PDG (2021) dataset, the electroweak fit yields a satisfactory result with \( \chi^2_{\text{min}}(\text{dof}) = 18.73(16) \), reflecting consistency across input observables. However, substituting the new CDF (2022) measurement sharply increases the minimum chi-squared to \( 64.45(16) \), revealing significant sensitivity in the fit—particularly in parameters such as \( m_W \), \( m_t \), \( m_Z \), and the hadronic contribution \( \Delta\alpha^{(5)}_{\text{had}} \). This divergence signals that the new data may point to potential physics beyond the Standard Model or unresolved experimental-systematic effects. In the context of electroweak symmetry breaking, the relation between the \(W\) and \(Z\) boson masses arises from the underlying SU(2)\(_L\) \(\times\) U(1)\(_Y\) gauge symmetry. At tree level, this relation is governed by the weak mixing angle \(\theta_W\), and can be approximated as:

\begin{equation}
m_W = m_Z \cos\theta_W,
\end{equation}
However, this simple form is modified by higher-order corrections, primarily due to the mass splitting between the top and bottom quarks, which introduces custodial symmetry breaking. This leads to a correction encoded in the \(\rho\) parameter:
\begin{equation}
\rho = 1 - \big[ \Pi_Z(0) - \Pi_W(0) \big] 
\equiv \frac{m_W^2}{m_Z^2 \cos^2\!\theta_W} 
\simeq 1 + \Delta\rho,
\end{equation}

where the running of \(\sin^2\theta_W(m_Z)\) and the precise top mass enter crucially. Precision electroweak fits constrain \(\sin^2\theta_\text{eff}\) with high accuracy, Consequently any observed discrepancy in \(m_W\) must be addressed through correlated shifts in \(m_t\) or \(m_Z\), highlighting the sensitivity of the weak scale to heavy virtual states.

Despite attempts to enhance \( m_W \), the discrepancy between the optimal values of \( m_Z \) and \( m_t \) remains approximately \(2\sigma\) from the initial parameters. As a result, the best-fit values still fail to reach the newly measured \( m_W \) reported by CDF. This discrepancy is clearly reflected in the significant negative pull shown in Fig.~\ref{fig:EWfit}. Additionally, it is important to highlight the \(2.8\,\sigma\) difference between the two most precise measurements of the top quark mass: \( 174.98 \pm 0.76 \, \text{GeV} \) from DØ~\cite{Abazov:2014fha} and \( 172.25 \pm 0.63 \, \text{GeV} \) from CMS~\cite{Sirunyan:2018koj}.

The global electroweak (EW) fit incorporating the CDF (2022) measurement of $m_W$ tends to favor a larger top-quark mass, reflecting its quadratic contribution to $\Delta \rho$ in

\begin{equation}
m_W^2 = \frac{m_Z^2}{2} \left[ 1 + \sqrt{1 - \frac{4 \pi \alpha}{\sqrt{2} G_\mu m_Z^2} (1 + \Delta r)} \,\right],
\end{equation}

where $\Delta r$ denotes the electroweak radiative corrections~\cite{Sirlin:1980nh,Hollik:1990ii}. In our numerical implementation, these corrections were evaluated using \texttt{FeynArts}~\cite{FeynArts} and \texttt{LoopTools}~\cite{LoopTools}. They can be written as

\begin{equation}
\Delta r = \Delta \alpha - \frac{c_W^2}{s_W^2}\,\Delta \rho + \Delta r_{\text{rem}},
\end{equation}

with $\Delta \rho \approx \dfrac{3 G_\mu m_t^2}{8 \sqrt{2} \pi^2}$ denoting the leading top-quark contribution and $\Delta r_{\text{rem}}$ summarizing subleading terms. Within the fit, the $W$-boson mass exhibits an anticorrelation with the hadronic contribution to the running of $\alpha$, $\Delta \alpha^{(5)}_{\text{had}}$, such that the elevated $m_W$ from CDF is associated with a reduced $\Delta \alpha^{(5)}_{\text{had}}$.

Moreover, this reduction in \( \Delta \alpha_{\text{had}} \) contributes to a smaller hadronic vacuum polarization correction to the muon anomalous magnetic moment \( a_{\mu}^{\text{HV}} \)	~\cite{deRafael:2020uif}. This in turn could increase the difference between the experimental \( a_{\mu}^{\text{Exp}} \) and the Standard Model prediction \( a_{\mu}^{\text{SM}} \), particularly when \( a_{\mu}^{\text{HV}} \) is derived from the global EW fit's \( \Delta \alpha_{\text{had}} \). Finally, the previous tension regarding the forward-backward asymmetry \( A_0, b_{\text{FB}} \) (or \( A_\ell \)) in the PDG (2021) results has become more pronounced (or alleviated) when analyzed within the context of the latest electroweak fits.

\section{Updated Constraints and Insights into New Physics}
\begin{figure}[!t]
	\centering
	\includegraphics[width=0.999\columnwidth]{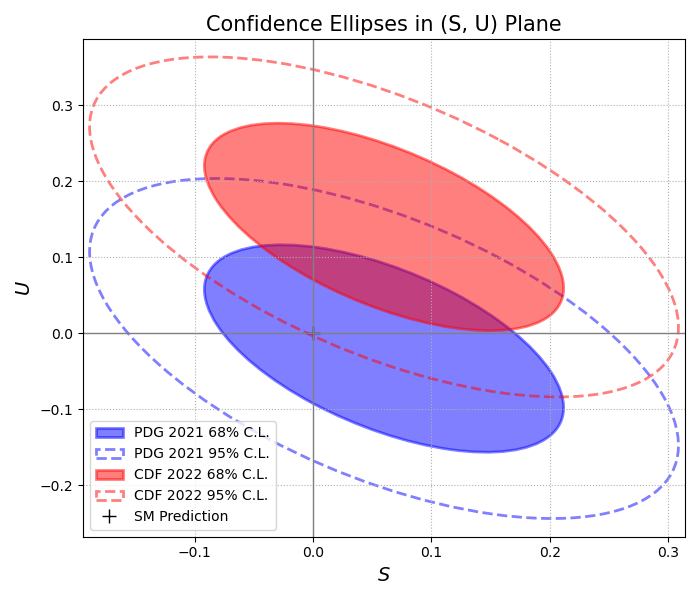}
	\caption{Allowed regions at $1\sigma$ and $2\sigma$ confidence levels in the $S$--$U$ plane, derived from electroweak fits. The blue contours correspond to the PDG 2021 dataset using the previous $m_W$ value, while the red contours include the updated CDF 2022 measurement. The inclusion of the CDF result leads to a visible positive shift in $U$, while $S$ remains approximately stable. The Standard Model prediction, corresponding to $(S,U) = (0,0)$, is indicated by the black cross at the origin.
	} \label{fig:SUfit}
\end{figure}

\begin{figure}[!t]
	\centering
	\includegraphics[width=\columnwidth]{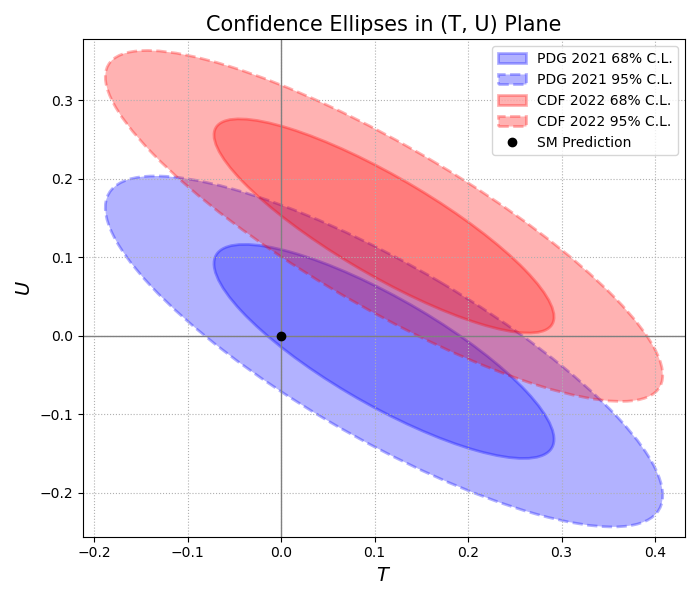}
	\caption{Allowed regions at $1\sigma$ and $2\sigma$ confidence levels in the $T$--$U$ plane, derived from electroweak fits using the global fit results of Table~\ref{tab:STUresult}. The blue contours correspond to the PDG 2021 dataset using the previous $m_W$ value, while the red contours include the updated CDF 2022 measurement. The inclusion of the CDF result leads to a positive shift in $U$, while $T$ remains essentially unchanged. The Standard Model prediction, corresponding to $(T,U) = (0,0)$, is indicated by the black dot at the origin.}
	\label{fig:TUfit}
\end{figure}

The electroweak precision fit results for the oblique parameters $S$, $T$, and $U$, including their correlations, are presented below. These results incorporate the $W$ boson mass measurements from CDF~(2022) and PDG~(2021), respectively. 
\begin{table}[ht]
	\centering
	\resizebox{\columnwidth}{!}{%
		\begin{tabular}{@{}lccc@{\hspace{1.5em}}ccc@{}}
			\toprule
			\multirow{2}{*}{13 dof} & \multicolumn{3}{c}{PDG 2021} & \multicolumn{3}{c}{CDF 2022} \\
			\cmidrule(lr){2-4} \cmidrule(lr){5-7}
			& Result & \multicolumn{2}{c}{Correlation} & Result & \multicolumn{2}{c}{Correlation} \\
			& $\chi^2_{\min}=15.42$ & S\quad T\quad U &  & $\chi^2_{\min}=15.44$ & S\quad T\quad U & \\
			\midrule
			S & $0.06 \pm 0.10$ & 1.00\quad0.90\quad$-0.57$ & & $0.06 \pm 0.10$ & 1.00\quad0.90\quad$-0.59$ & \\
			T & $0.11 \pm 0.12$ & \phantom{1.00\quad}1.00\quad$-0.82$ & & $0.11 \pm 0.12$ & \phantom{1.00\quad}1.00\quad$-0.85$ & \\
			U & $-0.02 \pm 0.09$ & \phantom{1.00\quad1.00\quad}1.00 & & $0.14 \pm 0.09$ & \phantom{1.00\quad1.00\quad}1.00 & \\
			\bottomrule
	\end{tabular}}
	\caption{Fit results for the oblique parameters $S$, $T$, and $U$ from the global electroweak fit .Correlation coefficients among parameters are also shown.}
	\label{tab:STUresult}	
\end{table}
The fits are performed using $m_h = 125\,\mathrm{GeV}$ and $m_t = 172.5\,\mathrm{GeV}$ as reference values for the Higgs boson and top quark masses. The data is adapted from Ref.~\cite{Lu:2022}
As shown in Table~\ref{tab:STUresult}, the values of $S$ and $T$ are presented under the assumption $\Delta U = 0$, following the same structure as Table~II. The results are taken from Ref.~\cite{Lu:2022}
\begin{table}[ht]
	\centering
	\resizebox{\columnwidth}{!}{%
		\begin{tabular}{@{}lc@{\hspace{1em}}c@{\hspace{0.5em}}c@{\hspace{1.5em}}c@{\hspace{0.5em}}c@{}}
			\toprule
			\multirow{2}{*}{\begin{tabular}[c]{@{}l@{}}$U = 0$\\14 dof\end{tabular}} 
			& \multicolumn{2}{c}{PDG 2021} & \multicolumn{2}{c}{CDF 2022} \\
			\cmidrule(lr){2-3} \cmidrule(lr){4-5}
			& Result & Correlation & Result & Correlation \\
			& $\chi^2_{\min}=15.48$ & S\quad T & $\chi^2_{\min}=17.82$ & S\quad T \\
			\midrule
			S & $0.05 \pm 0.08$ & 1.00\quad0.92 & $0.15 \pm 0.08$ & 1.00\quad0.93 \\
			T & $0.09 \pm 0.07$ & \phantom{1.00\quad}1.00 & $0.27 \pm 0.06$ & \phantom{1.00\quad}1.00 \\
			\bottomrule
		\end{tabular}%
	}
	\caption{Fit results for $S$ and $T$ parameters with $\Delta U = 0$ from PDG 2021 and CDF 2022.}
	\label{tab:STresult}
\end{table}
To illustrate the impact of the recent CDF 2022 measurement of the $W$-boson mass on the electroweak precision observables, we perform a global fit of the oblique parameters $S$ and $T$, assuming $\Delta U = 0$.
\begin{figure}[htbp]
	\centering
	\includegraphics[width=\columnwidth]{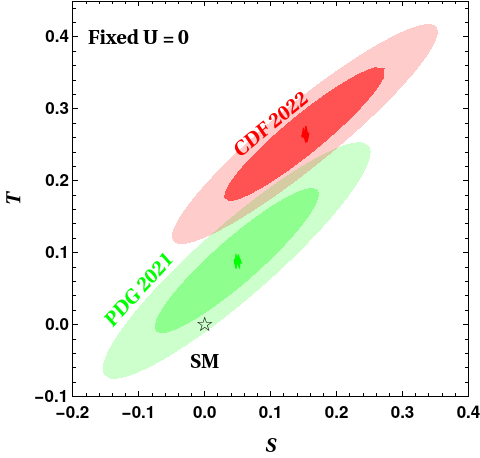} 
	\caption{Allowed regions at $1\sigma$ and $2\sigma$ confidence levels in the $S$--$T$ plane. The green contours correspond to the PDG 2021 dataset using the previous $m_W$ value, while the red contours include the updated CDF 2022 measurement. The observed upward shift in $T$ suggests possible new physics effects, consistent with our findings and those reported in ~\cite{Lu:2022}.}
	\label{fig:STfit}
\end{figure}
The corresponding allowed regions at $1\sigma$ and $2\sigma$ confidence levels are shown in Fig.~\ref{fig:STfit}. In addition to the oblique parameters $S$ and $T$, we also analyze the impact of the recent CDF 2022 measurement of the $W$-boson mass in the $S$--$U$ plane. As shown in Fig.~\ref{fig:SUfit}, the inclusion of the CDF result leads to a significant positive shift in $U$, while $S$ remains essentially unchanged compared to the determination based on the PDG 2021 dataset. The numerical inputs used in this fit are taken directly from ~\ref{tab:STUresult}, where the central values and uncertainties for $S$ and $U$, as well as their correlation coefficients, are employed to generate the confidence ellipses. Although the sensitivity of $U$ to new physics is generally weaker than that of $T$, the observed deviation in $U$ may still point to possible contributions from new electroweak multiplets or non-degenerate states that modify the self-energies of the charged and neutral gauge bosons.
Furthermore, it is instructive to examine the allowed regions in the $T$--$U$ plane within the framework of the Two-Higgs-Doublet Model (2HDM), as shown in Fig.~\ref{fig:TUfit}. This projection highlights the sensitivity of the electroweak fits to custodial symmetry breaking and isospin-violating effects induced by the extended scalar sector. The numerical inputs used to determine the shape of the confidence regions in this plane are taken directly from the global fit results presented in Table~\ref{tab:STUresult}, where the CDF 2022 dataset yields $T = 0.11 \pm 0.12$, $U = 0.14 \pm 0.09$, with a correlation coefficient $\rho_{TU} = -0.85$. Compared to the PDG 2021 fit, where $U$ was previously consistent with zero ($U = -0.02 \pm 0.09$), the inclusion of the CDF measurement leads to a moderate positive shift in $U$, while the $T$ parameter remains essentially unchanged.
This behavior remains compatible with the general expectations of the 2HDM, where radiative contributions from mass splittings among the additional scalar states dominantly affect $T$, while contributions to $U$ remain typically suppressed due to approximate custodial symmetry. The observed shift in $U$ may thus point to a small residual custodial symmetry breaking in the scalar sector.
The ellipse orientation reflects correlated oblique corrections from scalar loops. The PDG–CDF shift corresponds to a coherent deformation of the allowed electroweak-fit region.
\subsection{Electroweak Precision Constraints on the 2HDM from Oblique Parameters}
Electroweak precision observables (EWPOs) have long provided a powerful indirect probe of physics beyond the Standard Model (SM). Among these, the oblique parameters \(S\), \(T\), and \(U\) characterize new physics effects on gauge boson self-energies. In models such as the Two-Higgs-Doublet Model (2HDM), loop corrections from the extended scalar sector can induce significant contributions to these parameters, especially \(T\), which is highly sensitive to mass splittings between the charged and neutral scalars~\cite{Haber:2006,Chankowski1999}.
\begin{figure}[!t]
	\centering
	\includegraphics[width=0.52\textwidth]{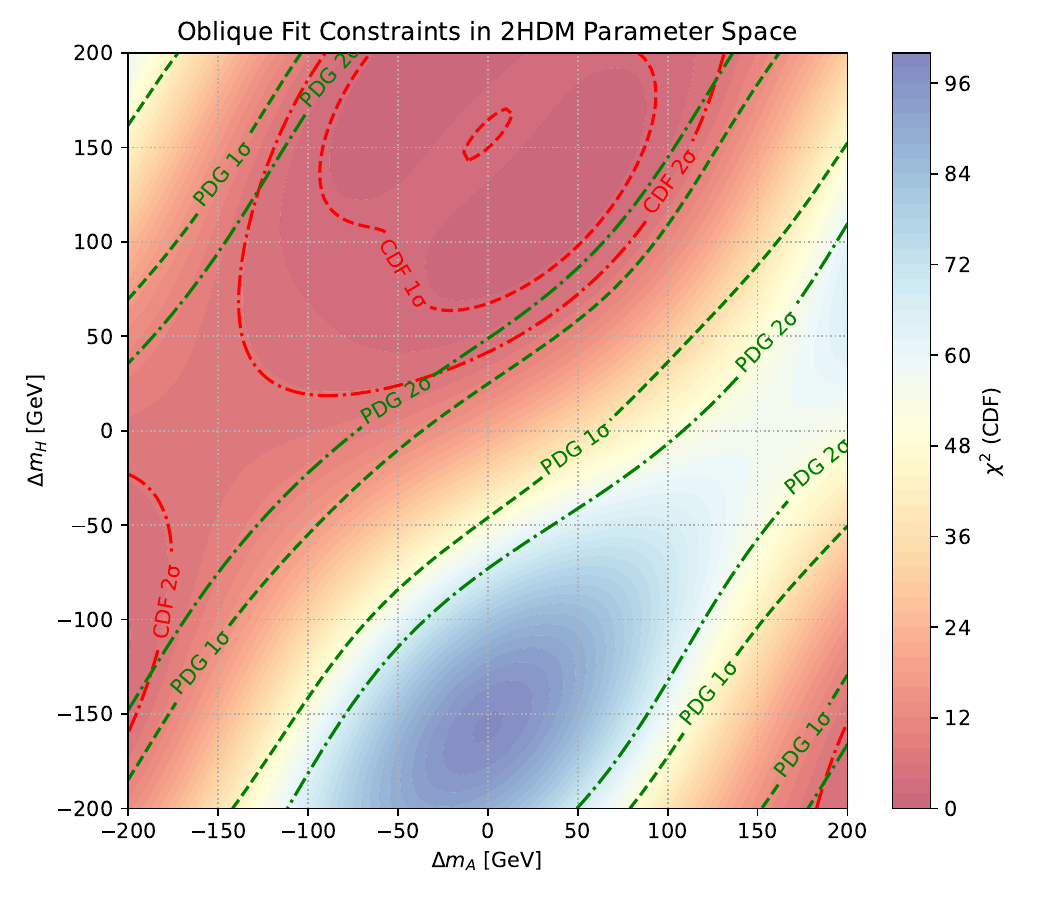}
	\caption{
		Contour map of the oblique parameters \( \Delta S \) and \( \Delta T \) in the Two-Higgs-Doublet Model (2HDM) as functions of scalar mass splittings \( \Delta m_A = m_A - m_h \) and \( \Delta m_H = m_H - m_h \), with \( m_h = 125\,\mathrm{GeV} \). The background color encodes the \(\chi^2\) deviation from the CDF 2022 electroweak fit~\cite{CDF:2022}. Overlaid are 1$\sigma$ and 2$\sigma$ confidence regions from CDF (red dashed and solid) and the PDG 2021 global fit (green dashed and solid). The alignment between theory and experiment highlights viable regions in the 2HDM parameter space that are consistent with electroweak precision tests and potentially indicative of new physics beyond the Standard Model.}
	\label{fig:2hdm-oblique}
\end{figure}
We focus on the case where \( U = 0 \), which is well motivated in many extensions of the SM, including custodial-symmetric scenarios. Figure~\ref{fig:2hdm-oblique} presents a contour map in the \((\Delta m_A,\, \Delta m_H)\) plane, where \( \Delta m_A = m_A - m_h \) and \( \Delta m_H = m_H - m_h \), with \( m_h = 125\,\mathrm{GeV} \). The heatmap reflects the \(\chi^2\) deviation from the best-fit point of the CDF 2022 electroweak fit~\cite{CDF:2022}, derived from the oblique parameters as in ~\ref{tab:STresult} \((S, T) = (0.15 \pm 0.08, 0.27 \pm 0.06)\) with correlation \( \rho = 0.93 \). 
\begin{figure}[!t]
	\centering
	\includegraphics[width=0.52\textwidth]{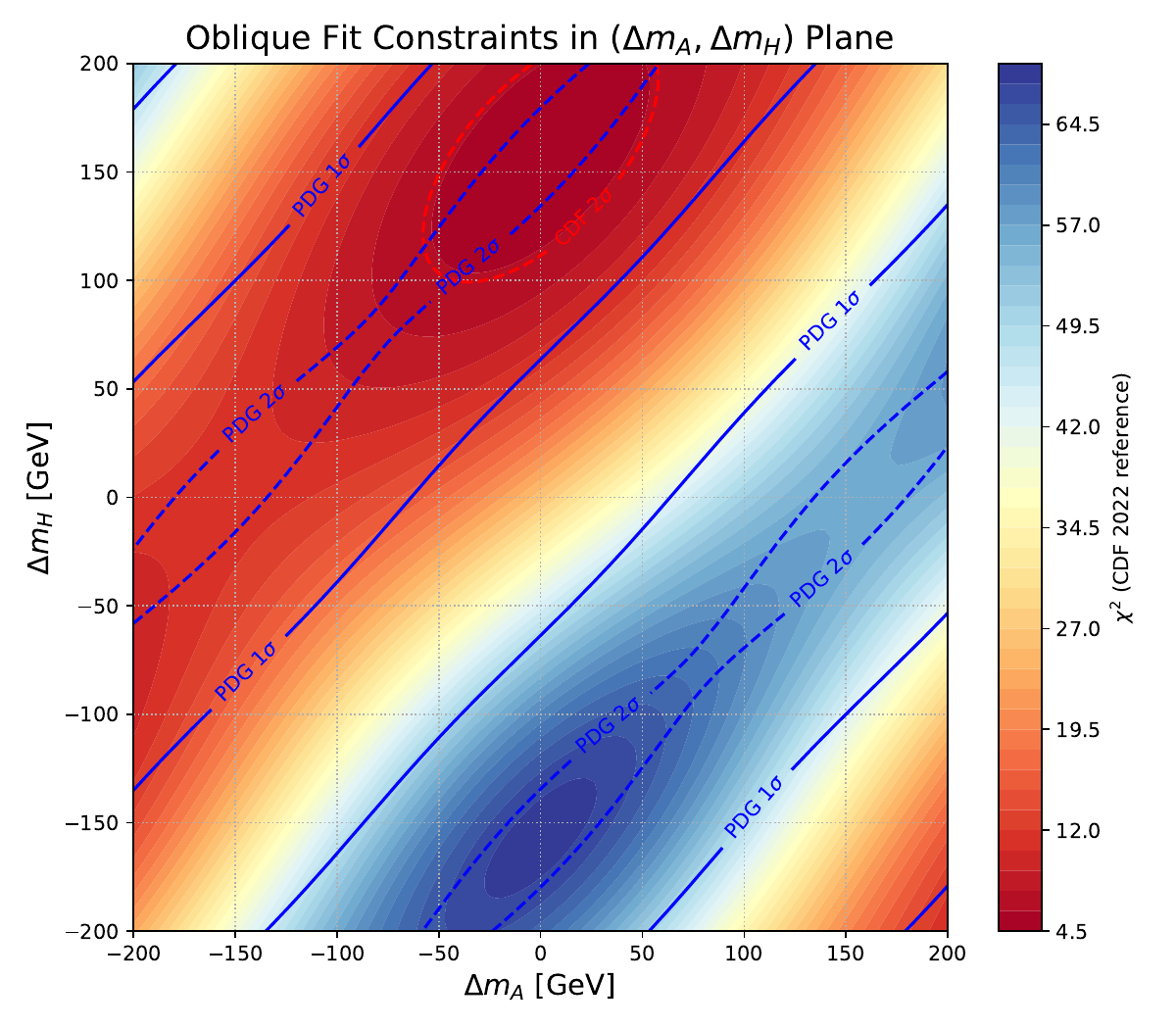}
	\caption{Allowed regions in the $(\Delta m_A, \Delta m_H)$ plane obtained from the oblique parameter fit in the Two-Higgs-Doublet Model. The colored background indicates the predicted $\chi^2$ values from the model calculations. The blue contours represent the $1\sigma$ and $2\sigma$ confidence levels derived from the PDG 2021 electroweak fit, while the red contours include the updated CDF 2022 measurement of the $W$ boson mass~\cite{CDF:2022}. The mass splittings $\Delta m_A$ and $\Delta m_H$ control the size of the custodial symmetry breaking effects entering the $S$ and $T$ parameters.
	}
	\label{fig:deltaMA_deltaMH}
\end{figure}
Overlaid are the 1$\sigma$ and 2$\sigma$ contours from both the CDF 2022 and PDG 2021 global electroweak fits. Notably, the region favored by the CDF measurement requires larger values of \(T\), which can be naturally achieved in the 2HDM via a mass hierarchy between the CP-even and CP-odd scalars. In contrast, the PDG-preferred region lies closer to the SM reference point. The intersection of the 2HDM predictions with these confidence regions provides a phenomenologically viable window for new physics, suggesting parameter regions testable at current or future colliders.
Furthermore, in order to investigate the sensitivity of the oblique parameters to variations in the scalar mass spectrum, we present in Fig.~\ref{fig:deltaMA_deltaMH} the allowed regions in the $(\Delta m_A, \Delta m_H)$ plane, where $\Delta m_A = m_A - m_{H^\pm}$ and $\Delta m_H = m_H - m_{H^\pm}$. These mass splittings directly affect the electroweak precision observables through loop contributions to the $S$ and $T$ parameters~\cite{Peskin:1990zt,Peskin:1991sw,Grimus:2008nb}. 
The colored background represents the predicted values of $\chi^2$ obtained from the Two-Higgs-Doublet Model (2HDM) Type-II as a function of the scalar mass splittings. Superimposed are the $1\sigma$ and $2\sigma$ contours derived from the global fits of the oblique parameters assuming $U=0$. The blue contours correspond to the fit using the PDG 2021 dataset, while the red contours represent the updated fit including the recent CDF 2022 measurement of the $W$-boson mass~\cite{CDF:2022}. 
As clearly seen, the inclusion of the CDF result shifts the allowed region towards larger values of $\Delta m_A$ and $\Delta m_H$, reflecting the preference for an enhancement in $T$ due to larger mass splittings within the scalar sector. This behavior is consistent with the expected contributions of non-degenerate scalar states in extended Higgs sectors such as the 2HDM~\cite{Haber:2010bw}. Overall, the fit results remain compatible with moderate mass splittings while still accommodating the observed $m_W$ anomaly.
\subsection{Probing Mass Splittings through the \texorpdfstring{$\Delta T$}{ΔT} Parameter}
Electroweak precision observables, and in particular the oblique parameter $\Delta T$, provide a sensitive probe of isospin-violating effects induced by physics beyond the Standard Model (BSM). The $\Delta T$ parameter quantifies deviations from custodial symmetry and is strongly affected by mass splittings within electroweak multiplets.
\begin{figure}[!t]
	\centering
	\includegraphics[width=\columnwidth]{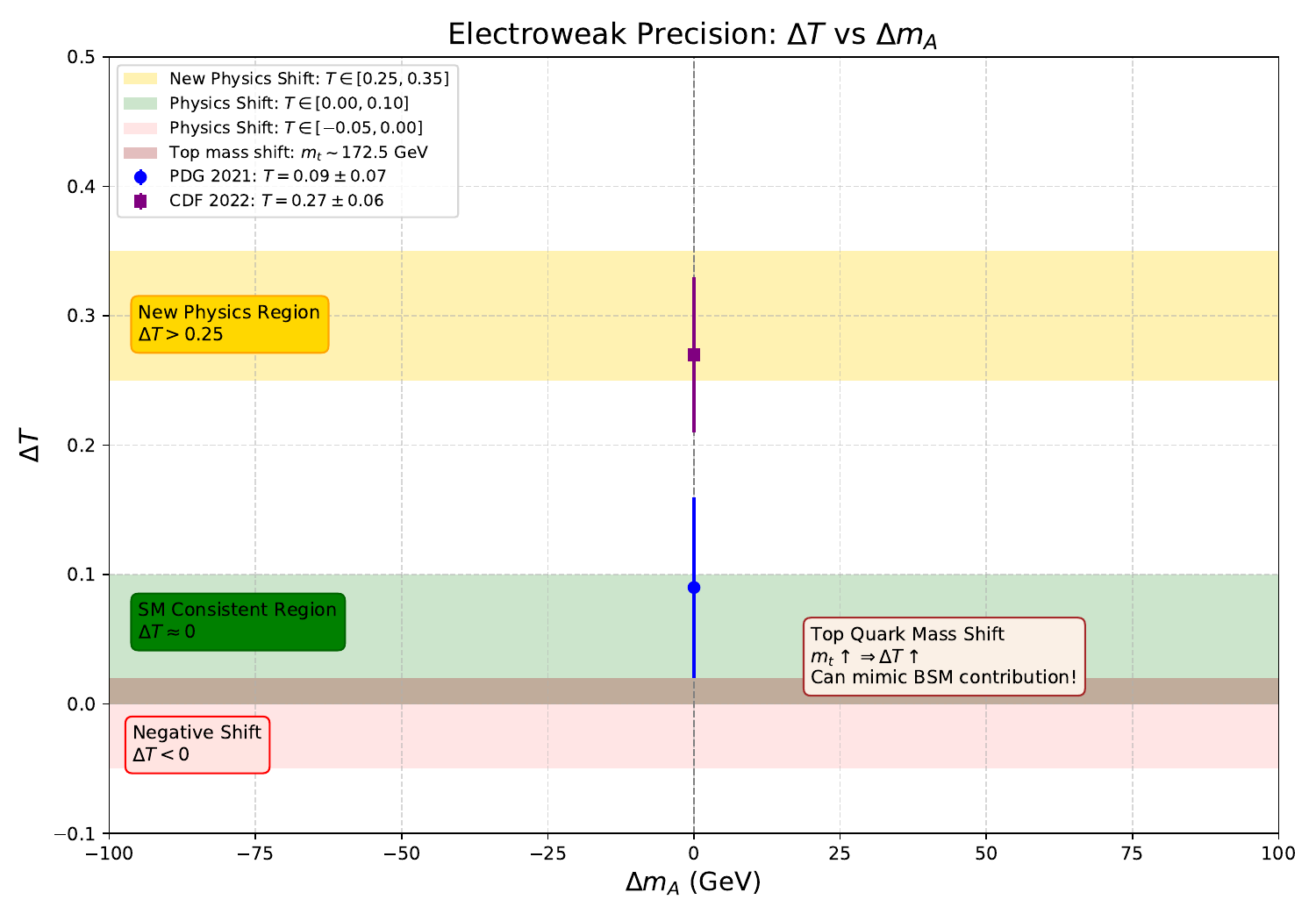}
	\caption{Correlation between the oblique parameter $\Delta T$ and the scalar mass splitting $\Delta m_A$. The horizontal shaded bands indicate representative regions: the Standard Model-compatible region ($0.00 \leq \Delta T \leq 0.10$, green), a region requiring additional positive contributions ($\Delta T > 0.25$, gold), and excluded negative values (red). The brown curve illustrates the contribution induced by variations of the top-quark mass around $m_t \sim 172.5~\mathrm{GeV}$. Experimental determinations from PDG 2021 and CDF 2022 are superimposed, highlighting the upward shift in $\Delta T$ suggested by the CDF $W$-boson mass measurement.
	}\label{fig:yourlabel}
\end{figure}
Such mass splittings naturally arise in extended scalar sectors, including Two-Higgs-Doublet Models (2HDMs), the Georgi--Machacek model, and models with vector-like fermions. A non-zero splitting between charged and neutral states generates a positive contribution to $\Delta T$, which can be directly compared to global electroweak fits.
The CDF measurement of the $W$ boson mass favors a larger value of $\Delta T$, indicating a possible departure from custodial symmetry. In this work, we parameterize this effect through $\Delta m_A$ as a representative scalar mass splitting and analyze its impact on $\Delta T$. By confronting these predictions with global fit results (PDG and CDF), we identify the regions of parameter space that remain consistent with electroweak precision constraints.
\begin{figure}[!t]
	\centering
	\includegraphics[width=\columnwidth]{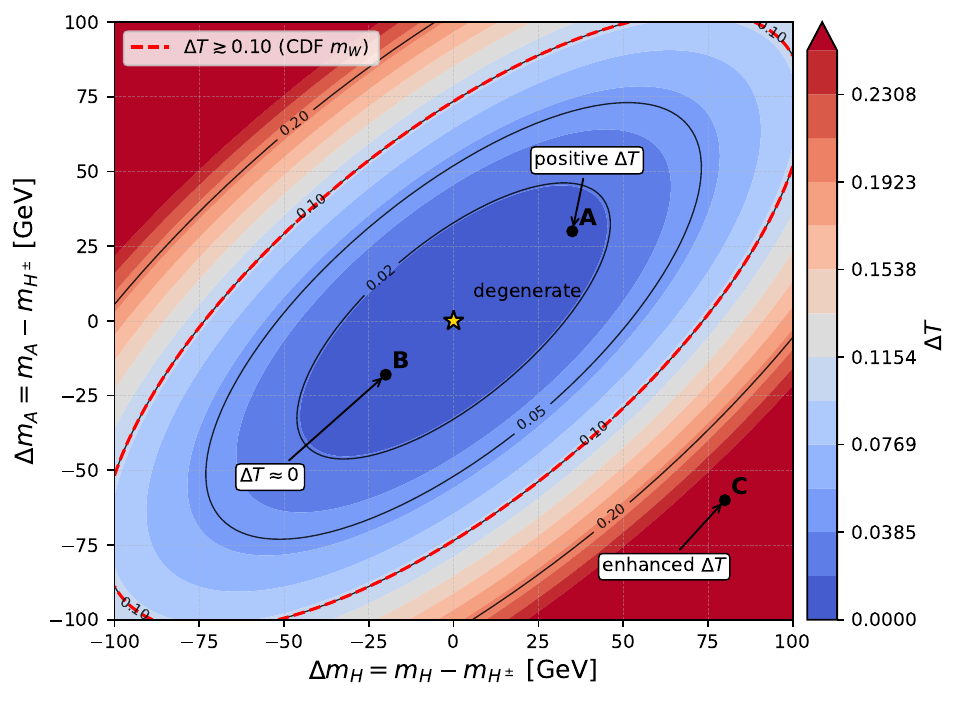}
	\caption{
		Scalar-mass-splitting contributions to the oblique parameter $\Delta T$ in the Type-II Two-Higgs-Doublet Model. The contour structure illustrates the sensitivity of $\Delta T$ to custodial-symmetry breaking induced by scalar mass splittings $\Delta m_H$ and $\Delta m_A$. The near-degenerate region corresponds to $\Delta T \approx 0$, while increasing mass splittings generate positive contributions to $\Delta T$. The dashed contour highlights the region $\Delta T \gtrsim 0.10$, which is relevant for accommodating the CDF $W$-boson mass anomaly within electroweak precision fits.
	}
	\label{fig:DeltaT_2HDM}
\end{figure}
To illustrate the physical interpretation of the parameter space, we highlight three representative benchmark points in the $(\Delta m_H,\, \Delta m_A)$ plane. Point B corresponds to the near-degenerate limit, $m_H \simeq m_A \simeq m_{H^\pm}$, where custodial symmetry is approximately preserved and $\Delta T \approx 0$. It therefore represents the reference configuration in which the extended scalar sector produces only a minimal correction
to electroweak precision observables. Point A represents a scenario with moderate scalar mass splittings, leading to a positive but controlled contribution to $\Delta T$. This region is of particular interest, as it can generate the required shift in electroweak precision observables suggested
by the CDF measurement of the $W$-boson mass. Point C illustrates a regime of large mass splittings, corresponding to strong custodial-symmetry breaking and
enhanced values of $\Delta T$. Such configurations are typically constrained by
electroweak precision data. These benchmark configurations demonstrate that the
relevant region for accommodating the CDF $m_W$ anomaly lies in an intermediate regime of controlled custodial-symmetry breaking, rather than in the extreme large-splitting limit. For definiteness, the mass splittings can be interpreted relative to a common charged Higgs mass scale,
$m_{H^\pm}$, which sets the overall scale of the scalar spectrum.

\subsection{Oblique Parameters in the 2HDM and Impact of the CDF-\texorpdfstring{$m_W$}{mW} Anomaly}

\begin{figure}[!t]
	\centering
	\includegraphics[width=\textwidth]{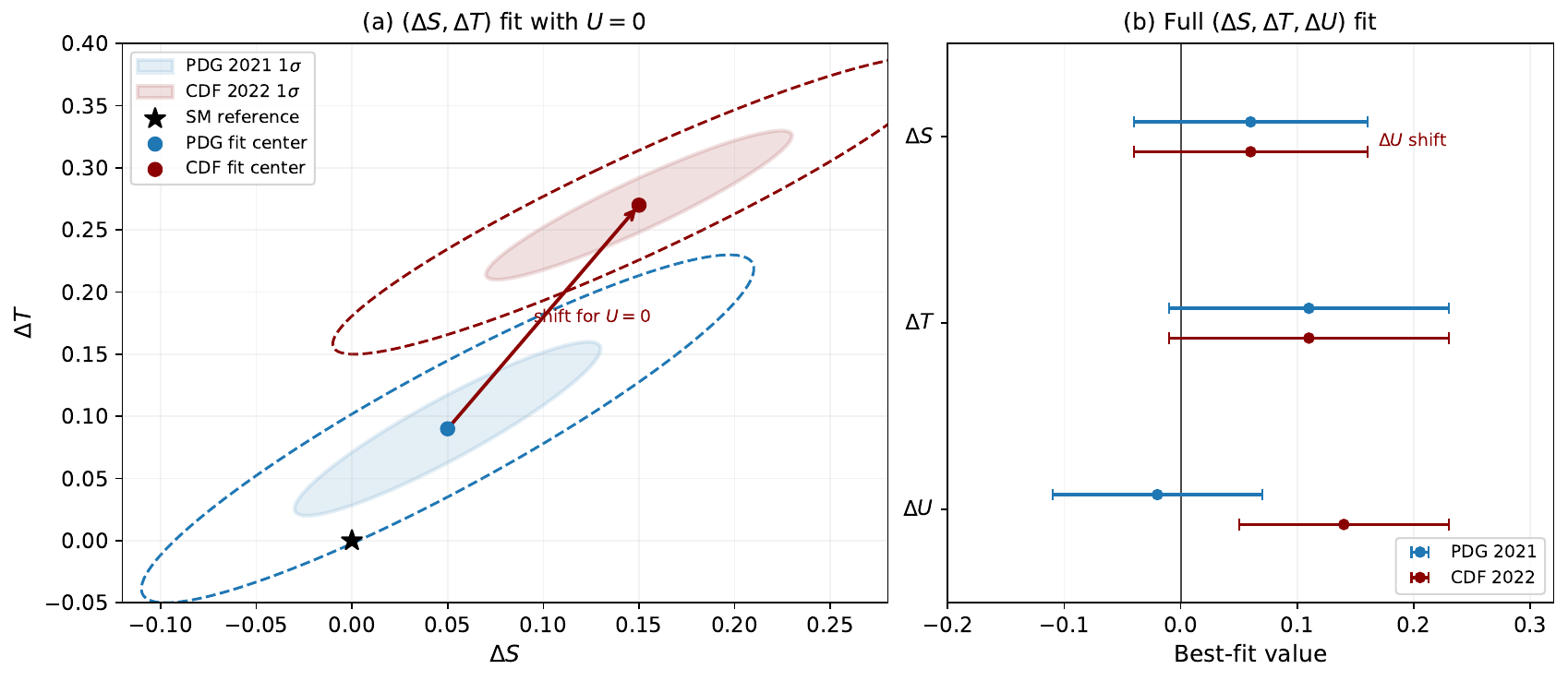}
	\caption{
		Electroweak precision constraints in the oblique parameter space. 
		(a) Confidence regions in the $(\Delta S,\Delta T)$ plane for the $U=0$ fit using PDG 2021 (blue) and CDF 2022 (red) inputs. 
		The shift between the two fit centers reflects the impact of the updated $W$-boson mass measurement, which favors larger values of $\Delta T$ under the $U=0$ constraint. 
		(b) Best-fit values of $(\Delta S,\Delta T,\Delta U)$ in the full STU analysis. 
		While $\Delta S$ and $\Delta T$ remain moderately stable, the CDF input induces a significant positive shift in $\Delta U$, indicating that the deviation is primarily absorbed along the $U$ direction when all oblique parameters are allowed to vary.
	}
	\label{fig:EWfit_STU}
\end{figure}

To assess the impact of the CDF measurement on electroweak precision observables, we compare global fits in the oblique parameter space. We consider both the constrained $U=0$ scenario and the full STU fit, allowing for a direct comparison of how the deviation is distributed among the oblique parameters. In contrast to illustrative parameter scans, the results shown here are directly constructed from global electroweak fit data, ensuring a model-independent interpretation of the observed shift.\\
As shown in Fig.~\ref{fig:EWfit_STU}, the $U=0$ constraint forces the deviation induced by the CDF measurement to be absorbed primarily in $\Delta T$, resulting in a visible displacement of the fit region in the $(\Delta S,\Delta T)$ plane. In contrast, the full STU fit reveals that the dominant effect is instead aligned with $\Delta U$, while $\Delta S$ and $\Delta T$ remain comparatively stable. This behavior highlights the importance of allowing all oblique parameters to vary when interpreting electroweak precision anomalies.\\
In the context of the 2HDM, this comparison is particularly instructive. A positive shift in $\Delta T$ can be generated by custodial-symmetry-breaking mass splittings in the scalar sector, whereas the full STU fit shows that part of the tension may instead be redistributed into $\Delta U$ once that degree of freedom is released. This implies that the interpretation of the CDF result within extended Higgs sectors is sensitive not only to the scalar spectrum itself, but also to the assumptions imposed on the electroweak fit. Consequently, the $U=0$ analysis provides a useful benchmark for identifying the scalar-mass patterns most relevant to the 2HDM, while the full STU result clarifies the extent to which the anomaly can be absorbed more generally in oblique corrections.

\clearpage
\onecolumngrid

\vspace{0.3cm}
\begin{figure}[!t]
	\centering
	\includegraphics[width=0.88\textwidth]{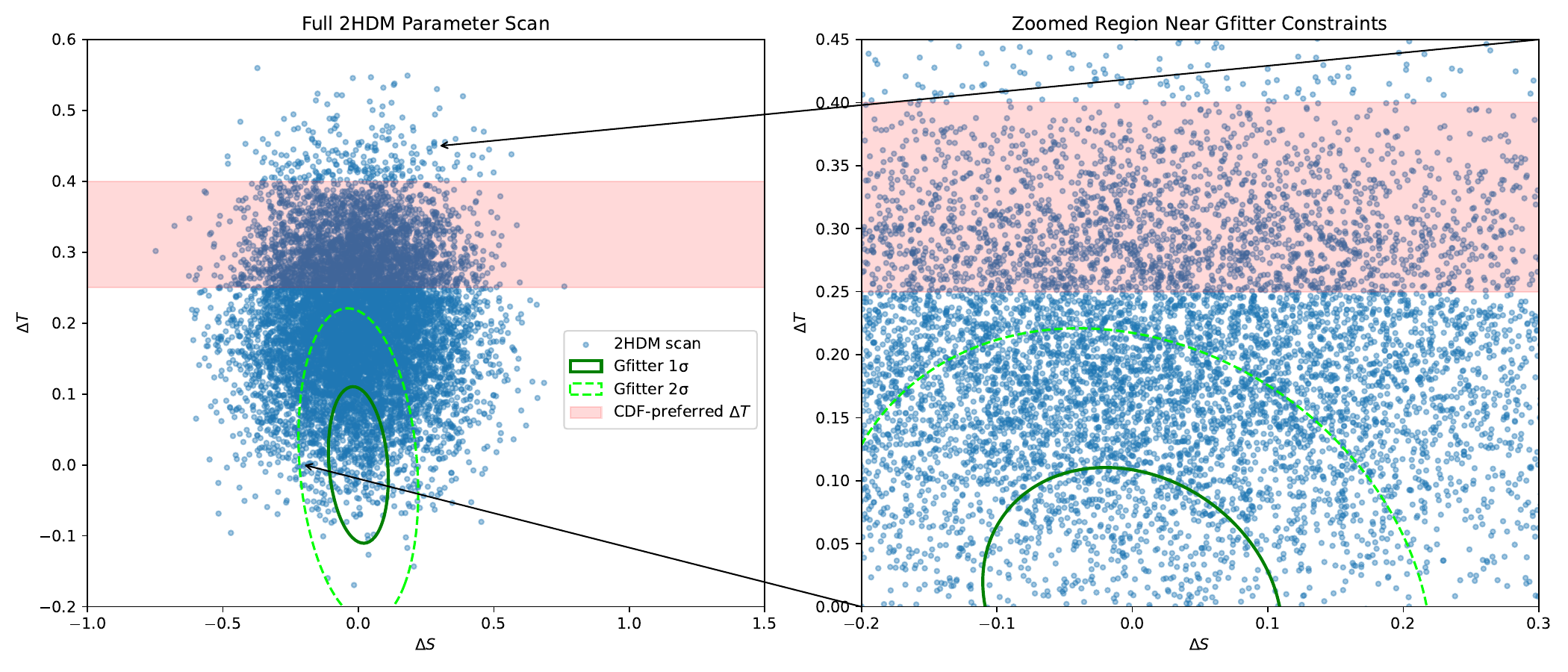}
	\caption{
		Oblique parameters in the 2HDM in the $(\Delta S,\Delta T)$ plane. 
		The scan points (blue) are compared with the electroweak-fit confidence regions from Gfitter~\cite{Baak:2014ora} and the CDF-preferred $\Delta T$ band (red) inferred from the measured shift in $m_W$~\cite{CDF:2022}. 
		The left panel shows the full parameter scan, while the right panel provides a zoom of the region compatible with precision constraints. 
		Only a restricted subset of points overlaps with the allowed region, indicating that viable configurations correspond to moderate scalar mass splittings that generate positive contributions to $\Delta T$ while keeping $\Delta S$ under control. 
		The scan was performed using \texttt{2HDMC}~\cite{Eriksson:2009ws}.
	}
	\label{fig:DeltaTvsDeltaS}
\end{figure}
\begin{minipage}[t]{0.48\textwidth}
	\noindent
	As illustrated in Fig.~\ref{fig:DeltaTvsDeltaS}, the 2HDM parameter scan populates a broad region in the $(\Delta S,\Delta T)$ plane, with only a subset overlapping the electroweak-fit contours~\cite{Baak:2014ora}. 
	This overlap becomes particularly relevant in light of the CDF II measurement of the $W$-boson mass~\cite{CDF:2022},
\end{minipage}
\hfill
\begin{minipage}[t]{0.48\textwidth}
	\noindent
	which favors a positive shift in $\Delta T$. 
	The results indicate that the accommodation of the CDF anomaly is not generic, but instead selects configurations with moderate scalar mass splittings that enhance $\Delta T$ while maintaining consistency with precision constraints.
\end{minipage}
\subsection{Global Fits in the \texorpdfstring{$(m_h,\, m_t)$}{(mh, mt)} Parameter Space from Electroweak Precision Data}

\begin{center}
	\includegraphics[width=0.48\textwidth]{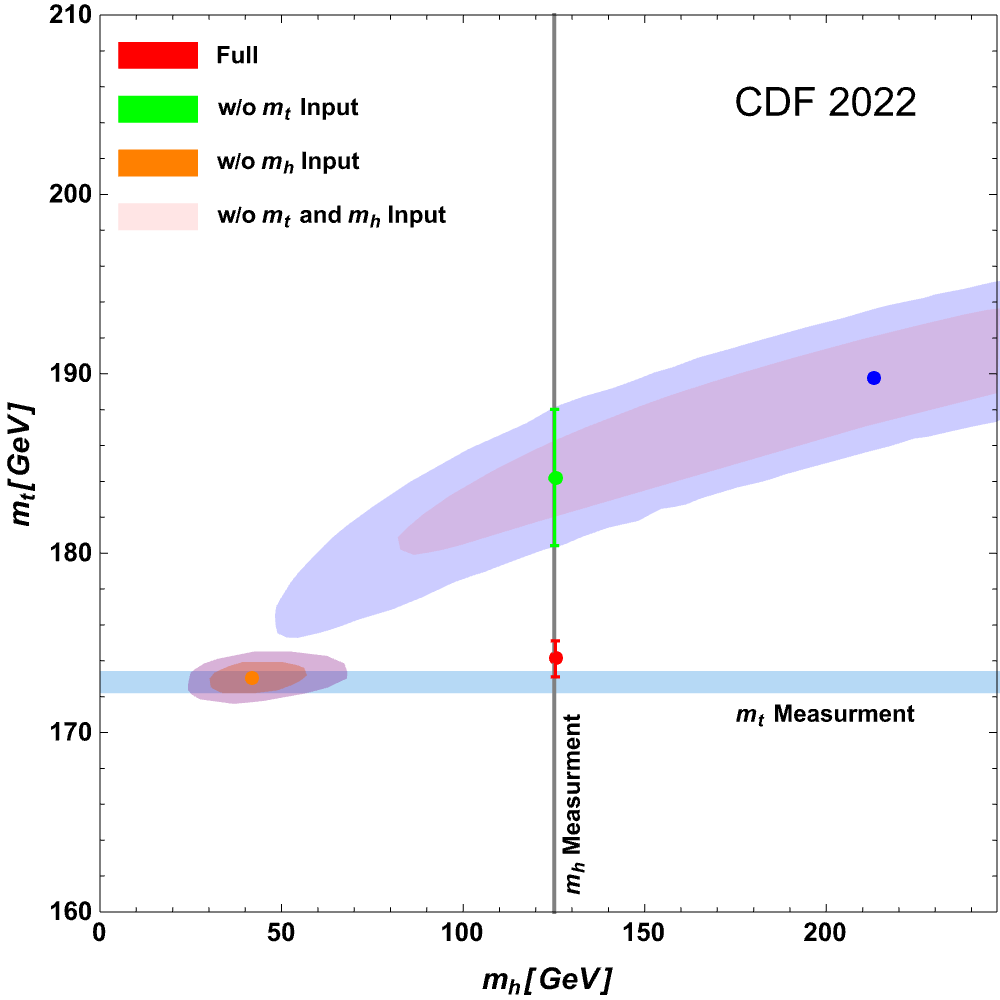}
	\hfill
	\includegraphics[width=0.48\textwidth]{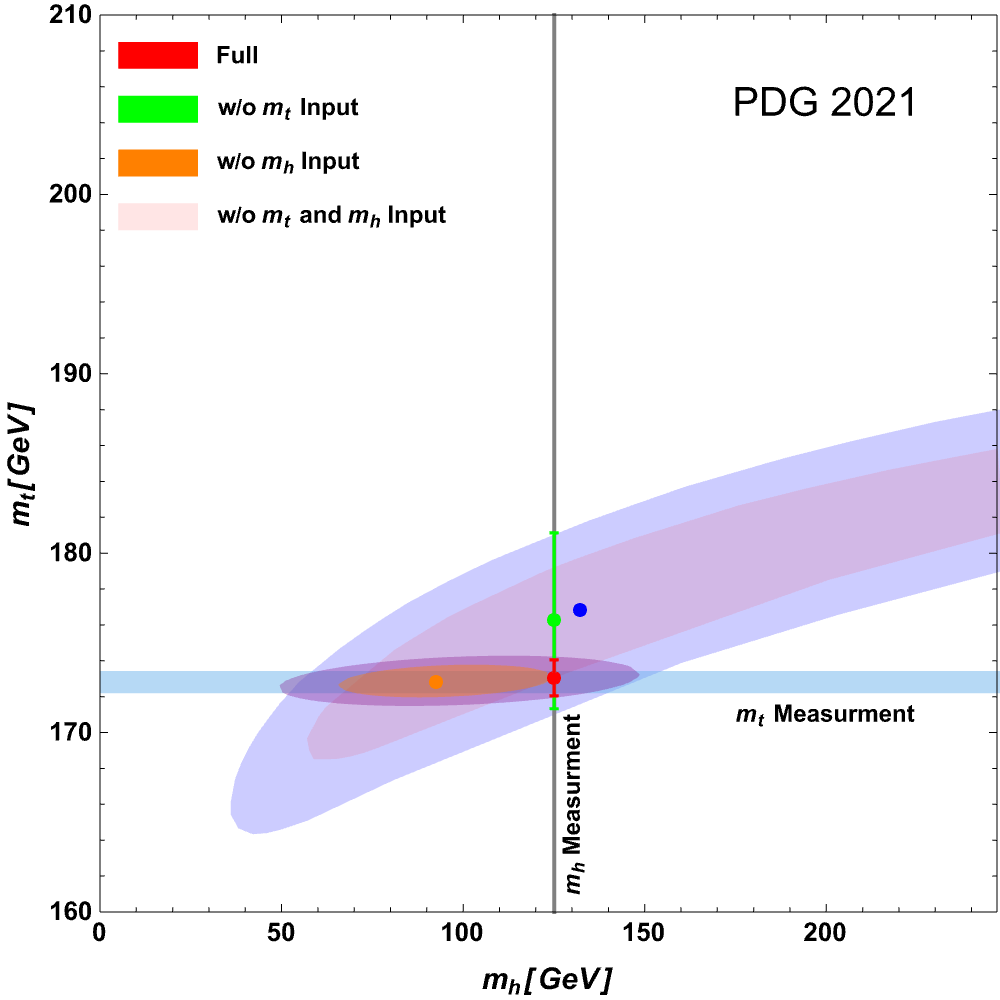}
\end{center}

\vspace{0.2cm}

\refstepcounter{figure}
\noindent
\textbf{Figure~\thefigure.}
Global electroweak fits in the $(m_h, m_t)$ plane.
\textbf{(a)} Fit including the CDF-II measurement of the $W$-boson mass. 
\textbf{(b)} Fit based on the PDG 2021 dataset. 
Colored regions correspond to different fit configurations: full fit (red), excluding $m_t$ (green), excluding $m_h$ (orange), and excluding both (pink). 
The horizontal blue and vertical grey bands indicate the direct experimental constraints on $m_t$ and $m_h$, respectively. 
The blue point denotes the CDF-II $m_W$ measurement. 
A visible shift of the preferred region toward larger $m_t$ values is observed in panel (a), highlighting the tension induced by the CDF result within the Standard Model electroweak fit.
\label{fig:fitmhmt_combined}
\vspace{0.2cm}
\twocolumngrid

\noindent
Figure~\ref{fig:fitmhmt_combined} shows the results of two-dimensional electroweak fits in the $(m_h,m_t)$ plane, consistent with Ref.~\cite{Lu:2022}. 
A strong correlation between the Higgs boson mass $m_h$ and the top-quark mass $m_t$ is observed, reflecting their joint role in electroweak radiative corrections.

Including the CDF-II $W$-boson mass measurement in panel~(a) shifts the preferred region toward larger values of $m_t$ compared to the PDG 2021 fit shown in panel~(b).
This behavior originates from the sensitivity of $m_W$ to heavy virtual states through loop corrections. 
In addition, removing the $m_t$ input (green contours) significantly enlarges the allowed region, highlighting its crucial role in constraining electroweak observables.

The resulting tension between the fitted and directly measured values of $m_t$,as reflected by the large positive pull shown in Fig.~\ref{fig:EWfit}, indicates that accommodating the CDF anomaly within the Standard Model is challenging and may point to contributions from new physics.

\section{Global \texorpdfstring{$\Delta \chi^2$}{Delta chi squared} Fits of Electroweak Masses}
\subsection{Top quark mass fit}

As the heaviest particle in the Standard Model (SM), the top quark plays a dominant role in electroweak radiative corrections. Its precise mass therefore provides a stringent test of the internal consistency of the global electroweak fit. Figure~\ref{fig:mt_delta_chi2} shows the $\Delta \chi^2$ profiles as functions of the top-quark mass $m_t$ for two global electroweak fits:

\begin{figure}[htbp]
	\centering
	\includegraphics[width=\columnwidth]{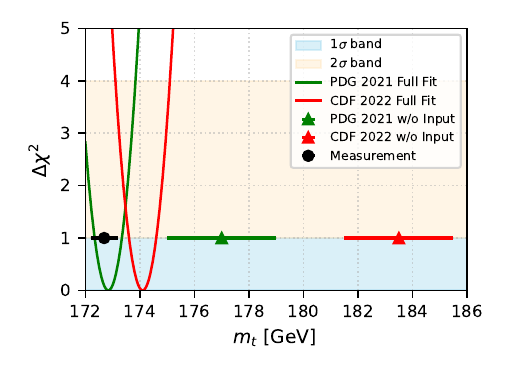}
	\caption{\label{fig:mt_delta_chi2}
		$\Delta \chi^2$ profiles as functions of the top-quark mass $m_t$ for the PDG 2021 and CDF 2022 global electroweak fits.
		The shaded regions denote the $1\sigma$ ($\Delta \chi^2 < 1$) and $2\sigma$ ($\Delta \chi^2 < 4$) confidence intervals.
		Triangular markers indicate fits excluding specific inputs, while the black circle shows the direct experimental measurement.}
\end{figure}
The Particle Data Group (PDG 2021) full fit (green) and the CDF 2022 full fit (red) exhibit a clear shift in the preferred value of $m_t$. The shaded regions indicate the $1\sigma$ and $2\sigma$ confidence intervals, reflecting the precision of the fit. Points with error bars correspond to fits excluding specific experimental inputs (denoted ``w/o Input''), illustrating the sensitivity of the electroweak fit to individual measurements. The black circular marker denotes the most recent direct determination of the top-quark mass. The shift observed in the CDF-based fit reflects the strong dependence of electroweak observables on loop corrections involving the top quark and indicates tension within the global fit.
\vspace{-15pt}
\subsection{Higgs Mass Determination from Global Electroweak Fits}
\begin{figure}[htbp]
	\centering
	\includegraphics[width=0.48\textwidth]{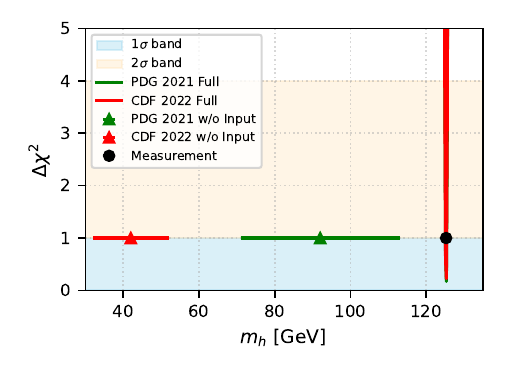}
	\caption{
		$\Delta \chi^2$ profiles for the Higgs boson mass $m_h$ obtained from global electroweak fits using PDG 2021 and CDF 2022 $m_W$ inputs. 
		Solid curves correspond to fits including all observables, while triangular markers denote fits excluding $m_h$ as an input. 
		The shaded bands indicate the $1\sigma$ (blue) and $2\sigma$ (orange) confidence regions. 
		A pronounced shift in the preferred $m_h$ value is observed in the CDF 2022 case, highlighting tension in the electroweak fit.
	}\label{fig:mh_fit_tension}
\end{figure}

\vspace{-7pt}
To assess the internal consistency of the global electroweak (EW) fit, we study the $\Delta \chi^2$ profile as a function of the Higgs boson mass $m_h$. This probes how strongly the remaining electroweak observables constrain $m_h$ when it is either included or excluded as a fit input. Figure~\ref{fig:mh_fit_tension} compares the resulting profiles for two
 scenarios: one based on the PDG 2021 world average of the $W$-boson mass and one incorporating the CDF 2022 result. In both cases, the solid curves correspond to fits in which $m_h$ is treated as a free parameter constrained by the remaining electroweak observables, while the triangular markers indicate values inferred when $m_h$ is excluded.
 
 A pronounced downward shift in the preferred Higgs mass is observed when $m_h$ is omitted, particularly in the CDF-based fit, where the minimum occurs near $m_h \sim 42~\mathrm{GeV}$, well below the measured value of $125.25~\mathrm{GeV}$. This discrepancy exceeds the $3\sigma$ level and signals a significant tension between indirect constraints and direct measurements. It indicates that accommodating the CDF $W$-boson mass within the Standard Model electroweak fit is nontrivial and may point to contributions beyond the Standard Model affecting electroweak precision observables.
 \vspace{-15pt}
\subsection{Electroweak Fit Constraints on \texorpdfstring{$m_Z$}{mZ}}
\vspace{-5pt}
\begin{figure}[htbp]
	\centering
	\includegraphics[width=0.98\columnwidth]{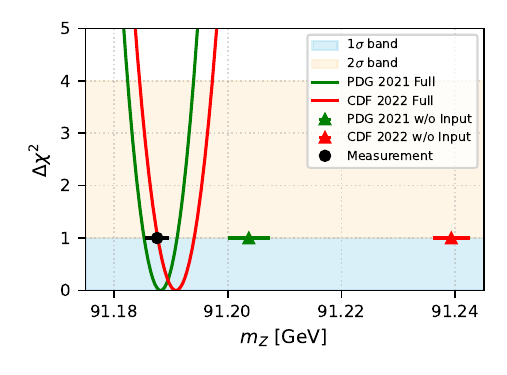}
	\caption{$\Delta \chi^2$ profiles for the $Z$ boson mass $m_Z$ obtained from electroweak fits using PDG 2021 and CDF 2022 datasets. 
		Solid curves correspond to global fits including $m_Z$ as an input, while markers with error bars denote best-fit values obtained when $m_Z$ is excluded. 
		The shaded regions indicate the $1\sigma$ (blue) and $2\sigma$ (yellow) confidence intervals. 
		The central black marker represents the direct LEP measurement of $m_Z$.	}
	\label{fig:mZchi2}
\end{figure}
Precise determination of the $Z$ boson mass, $m_Z$, is a cornerstone of electroweak precision tests of the Standard Model (SM), providing a stringent constraint on radiative corrections.

Figure~\ref{fig:mZchi2} shows the $\Delta \chi^2$ profiles as functions of $m_Z$ for both PDG 2021 and CDF 2022 datasets, distinguishing between full fits and fits excluding $m_Z$. The comparison between indirect determinations and the direct LEP measurement reveals the level of consistency within the electroweak fit.
In the CDF 2022 scenario, the fit excluding $m_Z$ favors a value shifted upward relative to the experimental average, indicating a non-negligible tension between indirect constraints and direct measurements. This shift reflects the sensitivity of electroweak observables to radiative corrections and suggests that accommodating the CDF $W$-boson mass may induce correlated distortions in other precision observables, including $m_Z$.
\vspace{-15pt}
\subsection{Probing Electroweak Precision via \texorpdfstring{$\Delta\alpha_{\text{had}}^{(5)}$}{Delta alpha had} Fits}
\begin{figure}[htbp] 
	\centering
	\includegraphics[width=\columnwidth]{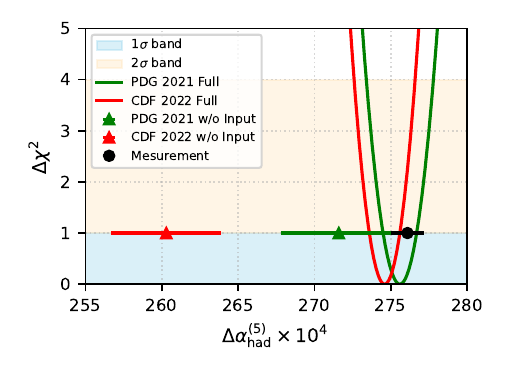}
	\caption{\label{fig:delta_alpha_chi2}
		$\Delta \chi^2$ profiles as functions of $\Delta\alpha_{\text{had}}^{(5)} \times 10^4$ for global electroweak fits based on PDG 2021 (green) and CDF 2022 (red) inputs. 
		The shaded bands denote the $1\sigma$ ($\Delta \chi^2 < 1$) and $2\sigma$ ($\Delta \chi^2 < 4$) confidence intervals. 
		Triangular markers indicate fits excluding specific inputs, illustrating the sensitivity to individual measurements. 
		The black circle denotes the input value used in the fits.
	}
\end{figure}

Figure~\ref{fig:delta_alpha_chi2} shows the $\Delta \chi^2$ dependence on the hadronic vacuum polarization contribution to the running of the
 electromagnetic coupling, $\Delta\alpha_{\text{had}}^{(5)}$, for both PDG 2021 and CDF 2022 electroweak fits. This quantity enters directly in the determination of the $W$-boson mass and other precision observables through radiative corrections.
The comparison between full and input-excluded fits highlights the sensitivity to $\Delta\alpha_{\text{had}}^{(5)}$. In the CDF-based fit, the preferred region shifts, indicating that accommodating the $W$-boson mass requires correlated modifications of $\Delta\alpha_{\text{had}}^{(5)}$, and thus points to potential beyond-SM contributions.
\vspace{-20pt}
\subsection{Precision Constraints from \texorpdfstring{$m_W$ and $\Delta \alpha^{(5)}_{\mathrm{had}}(M_Z)$}{mW and Delta Alpha}}
\vspace{-10pt}
Precision electroweak observables provide stringent tests of the Standard Model (SM) and sensitivity to new physics. The $W$-boson mass $m_W$ and the hadronic contribution $\Delta \alpha^{(5)}_{\mathrm{had}}(M_Z)$ are
\begin{figure}[htbp]
	\centering
	\includegraphics[width=0.99\linewidth]{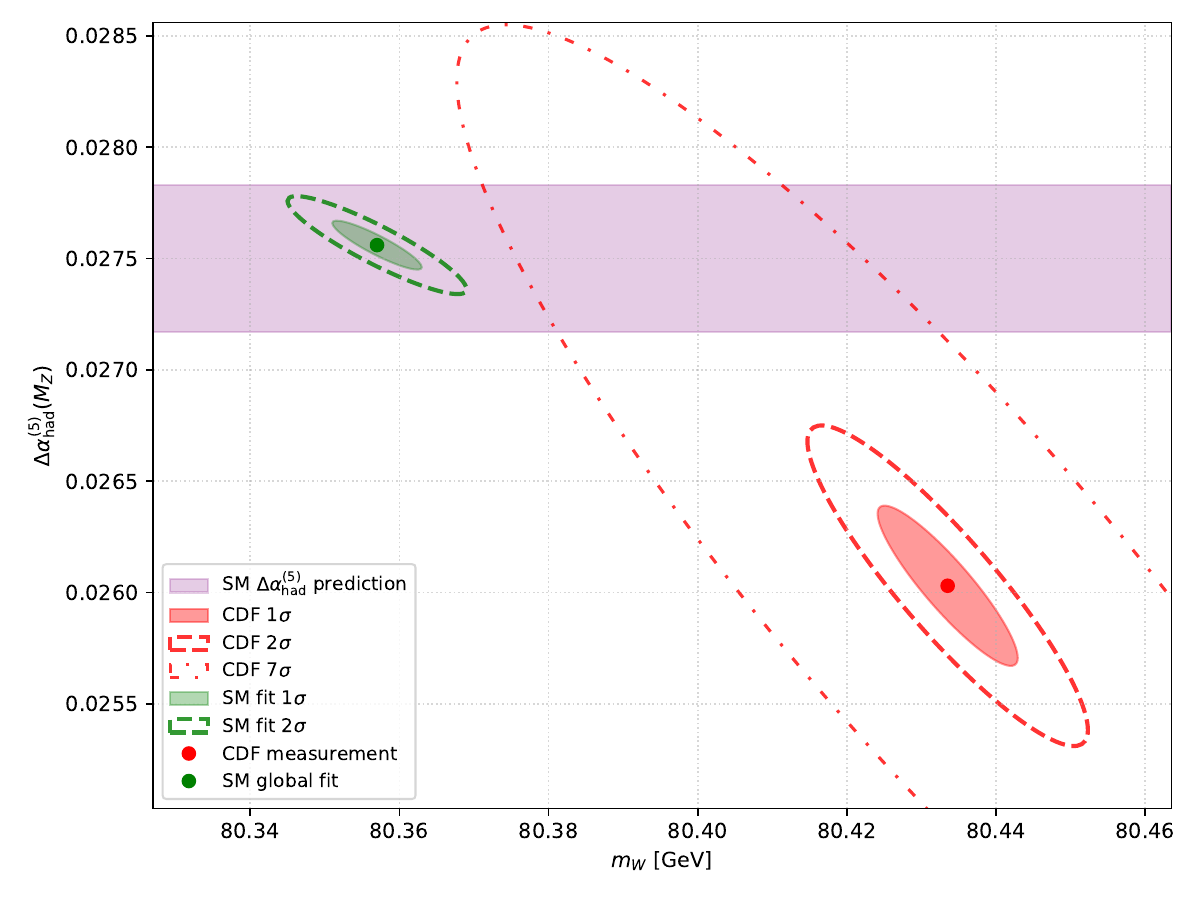}
	\caption{Confidence regions (1$\sigma$, 2$\sigma$, and 7$\sigma$) in the $(m_W, \Delta \alpha^{(5)}_{\mathrm{had}}(M_Z))$ plane, comparing the CDF measurement~\cite{CDF:2022} (red ellipses) with the SM global fit $m_W^{\text{SM}} = 80.357 \pm 0.006~\mathrm{GeV}$~\cite{Zyla:2020zbs} (green ellipses). Shaded vertical and horizontal bands represent the SM predictions and uncertainties for $m_W$ and $\Delta \alpha^{(5)}_{\mathrm{had}}$, respectively. 
		A strong negative correlation ($\rho = -0.9$) is assumed. 
		The separation between the CDF result and the SM prediction approaches $7\sigma$, indicating a significant tension.
		Figure~\ref{fig:ellipses} illustrates the global tension between the CDF measurement and the SM prediction, highlighting the magnitude of the deviation in the $(m_W, \Delta \alpha^{(5)}_{\mathrm{had}})$ plane.
	}
	\label{fig:ellipses}
\end{figure}
 particularly sensitive to loop effects, making deviations in $m_W$ a direct probe of radiative corrections and potential beyond-SM contributions~\cite{Awramik:2003rn,Freitas:2014hra,Degrassi:1997iy}.
Figure~\ref{fig:ellipses} shows that the CDF result lies significantly above the SM global fit, with a deviation approaching $7\sigma$. This tension reflects the strong interplay between $m_W$ and $\Delta \alpha^{(5)}_{\mathrm{had}}(M_Z)$ in the electroweak fit and suggests that accommodating the CDF measurement requires nontrivial modifications of the SM radiative structure.

\begin{figure}[htbp]
	\centering
	\includegraphics[width=0.98\linewidth]{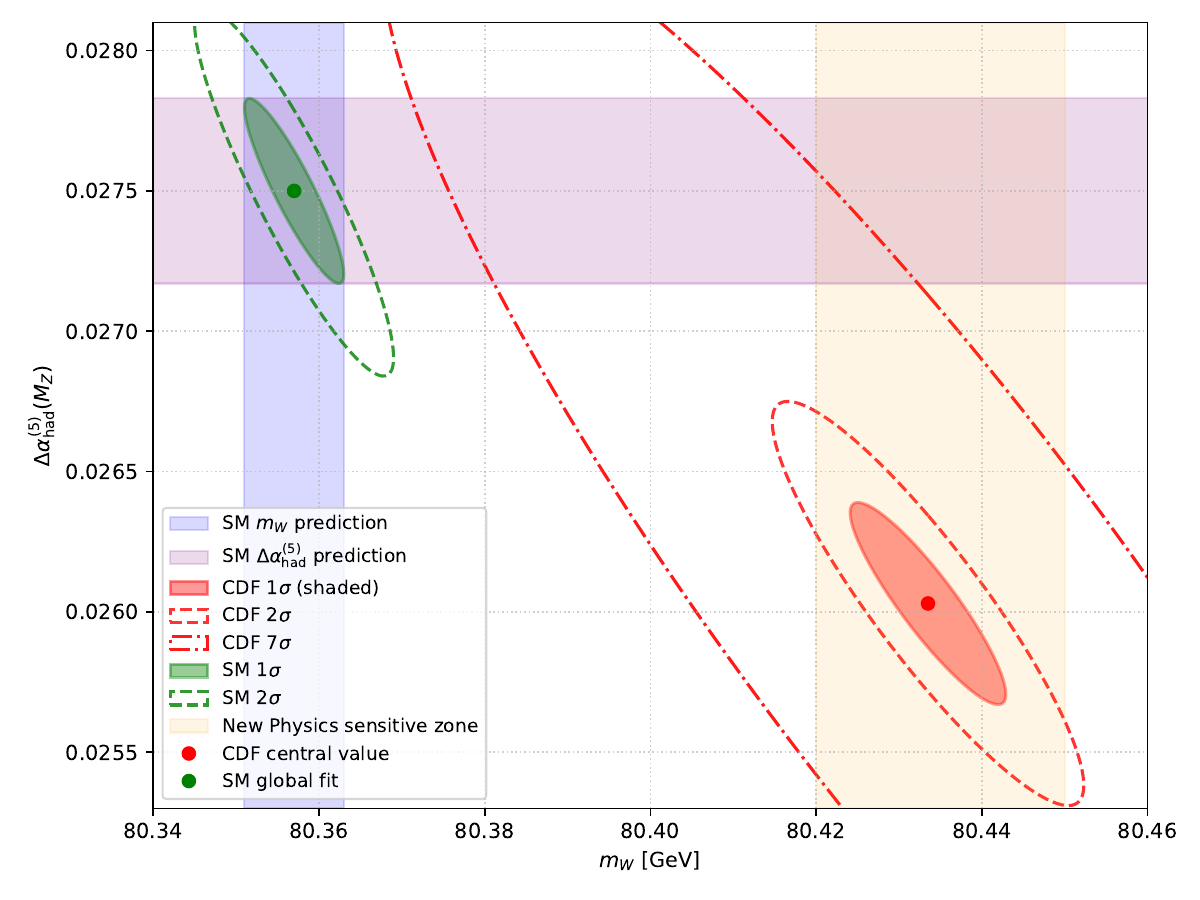}
	\caption{
		Elliptical confidence regions in the $(m_W, \Delta \alpha^{(5)}_{\text{had}}(M_Z))$ plane showing $1\sigma$ (solid) and $2\sigma$ (dashed) contours for the CDF II measurement (red) and the SM global fit (PDG 2021) (green). 
		A $7\sigma$ deviation contour is shown as a dash-dotted red ellipse. 
		Shaded bands indicate the SM predictions for $m_W$ and $\Delta \alpha^{(5)}_{\text{had}}$. 
		The highlighted region illustrates parameter space sensitive to new physics contributions.
		Figure~\ref{fig:mw-dalpha-ellipses} provides a refined view of the parameter space, identifying the region where new physics contributions can reconcile the observed deviation. 
	}
	\label{fig:mw-dalpha-ellipses}
\end{figure}

Such a deviation can be consistently interpreted in terms of oblique corrections, parameterized by the $S$, $T$, and $U$ parameters~\cite{Peskin:1990zt,Peskin:1991sw}. The magnitude of the shift points to new electroweak-scale contributions, as realized in extensions of the SM with additional scalar sectors, electroweak multiplets, or modified gauge dynamics~\cite{Altarelli:1991fk,deBlas:2022hdk}. Improving the determination of $\Delta \alpha^{(5)}_{\mathrm{had}}(M_Z)$—dominated by low-energy hadronic effects—is therefore essential for stabilizing the global electroweak fit and sharpening its sensitivity to new physics~\cite{deRafael:2020uif}
The structure of the confidence regions reflects the strong correlation between $m_W$ and $\Delta\alpha_{\mathrm{had}}^{(5)}(M_Z)$, both entering electroweak precision observables through radiative corrections. The CDF measurement shifts the preferred region relative to the SM global fit, indicating a coherent displacement in parameter space. This behavior points to modified loop contributions affecting the electroweak fit. The separation between the two regions therefore provides a sensitive probe of potential deviations from Standard Model expectations.
\subsection{Electroweak Precision Observables and the Effective Weak Mixing Angle}

Electroweak precision observables (EWPOs) provide a stringent and internally consistent framework for testing the Standard Model (SM) at the loop level~\cite{Schael2006,Baak:2014ora,DeBlas:2019okz}. Among these, the effective weak mixing angle $\sin^{2}\bar{\theta}_W$ plays a central role, encoding the interplay between neutral gauge boson mixing and electroweak radiative corrections.

\begin{figure}[htbp]
	\centering
	\includegraphics[width=1.02\linewidth]{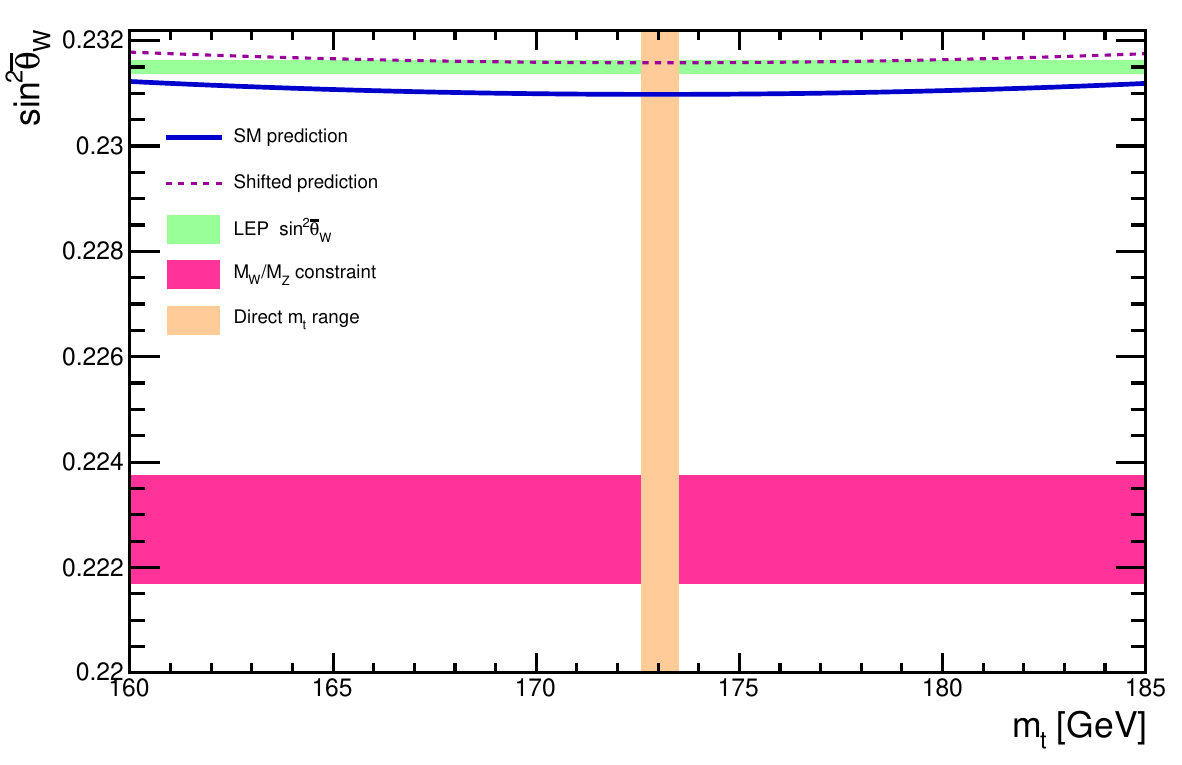}
	\caption{
		Dependence of the effective weak mixing angle $\sin^2 \bar{\theta}_W$ on the top-quark mass $m_t$. 
		The solid blue curve shows the Standard Model prediction including electroweak radiative corrections. 
		The dashed curve represents an illustrative shifted prediction induced by additional contributions to gauge-boson self-energies, such as a positive shift in the oblique parameter $\Delta T$. 
		The green band denotes the LEP determination of $\sin^2 \bar{\theta}_W$, the vertical orange band indicates the direct experimental range of $m_t$, and the horizontal shaded band indicates the constraint derived from the $M_W/M_Z$ relation. 
		The comparison highlights the correlation among electroweak precision observables and their sensitivity to loop-level effects.
	}
	\label{fig:s2w_mt}
\end{figure}

Within a model-independent framework, deviations in electroweak precision observables can be interpreted as loop-induced modifications of gauge-boson self-energies, parameterized by the oblique parameters $S$, $T$, and $U$~\cite{Peskin:1990zt,Peskin:1991sw}. These parameters provide a universal description of new physics effects that primarily affect vacuum polarization amplitudes.

In the SM, the dependence of $\sin^{2}\bar{\theta}_W$ on the top-quark mass arises from electroweak radiative corrections, exhibiting a quadratic sensitivity to $m_t$ and a logarithmic dependence on the Higgs boson mass. Consequently, the combined analysis of $\sin^{2}\bar{\theta}_W$, $m_W$, and $m_t$ constitutes a stringent test of electroweak consistency.

Figure~\ref{fig:s2w_mt} illustrates this correlation structure. The SM prediction (solid curve) provides a consistent description of the measured observables, while the dashed curve shows the impact of additional contributions to electroweak radiative corrections. In particular, a positive contribution to $\Delta T$ modifies the relation between $m_W$, $m_Z$, and $\sin^{2}\bar{\theta}_W$, inducing a correlated upward shift in the predicted value of $\sin^{2}\bar{\theta}_W$.

Such correlated shifts are directly relevant in light of the CDF measurement of the $W$-boson mass, which favors a higher value of $m_W$ compared to the SM prediction. In this context, the shifted curve provides a qualitative representation of how oblique corrections can alleviate tensions among electroweak precision observables.

These results demonstrate that even percent-level shifts in $\sin^{2}\bar{\theta}_W$ can carry significant implications for the structure of the electroweak sector, making precision measurements of this observable a sensitive probe of physics beyond the Standard Model.
\section{Conclusion}

In this work, we have investigated the implications of the CDF II measurement of the $W$-boson mass within the framework of electroweak precision observables and the Two-Higgs-Doublet Model (2HDM). 
The inclusion of the CDF result in the global electroweak fit leads to a substantial increase in the minimum chi-squared, from $\chi^2_{\mathrm{min}} = 18.73$ (PDG 2021) to $\chi^2_{\mathrm{min}} = 64.45$, indicating a significant tension with the Standard Model. This tension is not confined to $m_W$ alone, but propagates to other precision observables, including $m_t$, $m_Z$, and $\Delta \alpha_{\mathrm{had}}^{(5)}$, reflecting the strongly correlated structure of electroweak radiative corrections.
Within a model-independent framework, this discrepancy can be interpreted in terms of oblique corrections. In particular, the CDF measurement favors a sizable positive shift in the $T$ parameter, with
\begin{equation}
\Delta T \sim \mathcal{O}(0.2),
\end{equation}
while $S$ remains moderately affected and $U$ plays a subleading role. This pattern is consistent with new physics contributions that break custodial symmetry.
In the context of the 2HDM, such a positive contribution to $\Delta T$ can be generated through mass splittings among the scalar states. Our analysis shows that moderate mass hierarchies in the scalar sector can produce the required enhancement in $\Delta T$, thereby partially alleviating the tension induced by the CDF measurement. In particular, configurations with non-degenerate charged and neutral Higgs bosons lead to the largest contributions.
However, the global fit results indicate that the tension cannot be completely resolved within minimal extensions of the scalar sector alone. The required shift in $\Delta T$ is relatively large and may be subject to additional theoretical and experimental constraints, including perturbativity, vacuum stability, and direct collider bounds on scalar masses.
Overall, our results highlight the sensitivity of electroweak precision observables to loop-level effects and demonstrate that even small deviations in $m_W$ can have far-reaching implications for physics beyond the Standard Model. The observed pattern of deviations strongly points toward new contributions to gauge-boson self-energies, making precision electroweak fits a powerful tool for probing extended Higgs sectors and other scenarios of new physics.
Future measurements of the $W$-boson mass and improved determinations of electroweak observables will be crucial in clarifying the origin of the current tension and in further constraining viable extensions of the Standard Model.
\section*{Acknowledgements}

The authors sincerely thank Prof.~Driss Khalil and Prof.~Larbi Rahili. This research was performed using the MARWAN High-Performance Computing platform provided by the Moroccan National Center for Scientific and Technical Research (CNRST).

\subsection{Passarino--Veltman Functions}

The scalar two-point Passarino--Veltman functions entering the electroweak precision analysis are defined as follows.

\subsubsection*{Function $B_{22}$}

\begin{widetext}
	\begin{equation}
	B_{22}(q^2; m_1^2, m_2^2)
	= \frac{1}{2} q^2 \left( \Delta - \log\frac{\mu^2}{m_1 m_2} \right) + \kappa
	\end{equation}
\end{widetext}

where $\Delta$ denotes the divergent contribution and $\mu$ is the renormalization scale. The finite term $\kappa$ is

\begin{widetext}
	\begin{equation}
	\begin{aligned}
	\kappa ={}&
	\frac{1}{2} q^2 \log \frac{m_1 m_2}{\mu^2}
	-\frac{1}{2}(m_1^2+m_2^2-q^2) \\
	&\times
	\log\left(
	\frac{m_1^2+m_2^2-q^2+\sqrt{(m_1^2+m_2^2-q^2)^2-4m_1^2m_2^2}}
	{2m_1m_2}
	\right).
	\end{aligned}
	\end{equation}
\end{widetext}
\subsubsection*{Function $B_0$}
\begin{widetext}
	\begin{equation}
	B_0(q^2; m_1^2, m_2^2)
	= \Delta - \log \frac{\mu^2}{m_1 m_2} + \kappa
	\end{equation}
\end{widetext}

\subsection{Finite Contributions}
\begin{widetext}
	\begin{equation}
	\begin{aligned}
	\kappa ={}&
	-\frac{1}{2}\left(
	\log \frac{m_1^2}{\mu^2}
	+
	\log \frac{m_2^2}{\mu^2}
	\right) \\
	&+
	\frac{1}{2}
	\left(
	\frac{m_1^2+m_2^2-q^2}{q^2}
	\right)
	\log\left(
	\frac{m_1^2+m_2^2-q^2+\sqrt{(m_1^2+m_2^2-q^2)^2-4m_1^2m_2^2}}
	{2m_1m_2}
	\right).
	\end{aligned}
	\label{eq:kappa_B0}
	\end{equation}
\end{widetext}

These finite terms arise from loop integrals and enter directly into the oblique-parameter analysis.
\onecolumngrid
\section{Extended Electroweak Fit Diagnostics}
For completeness, we present an additional comparison of pull values between the PDG 2021 and CDF 2022 electroweak fits, highlighting the pattern of deviations across precision observables.
\begin{figure}[p]
	\centering
	\includegraphics[width=0.76\textwidth]{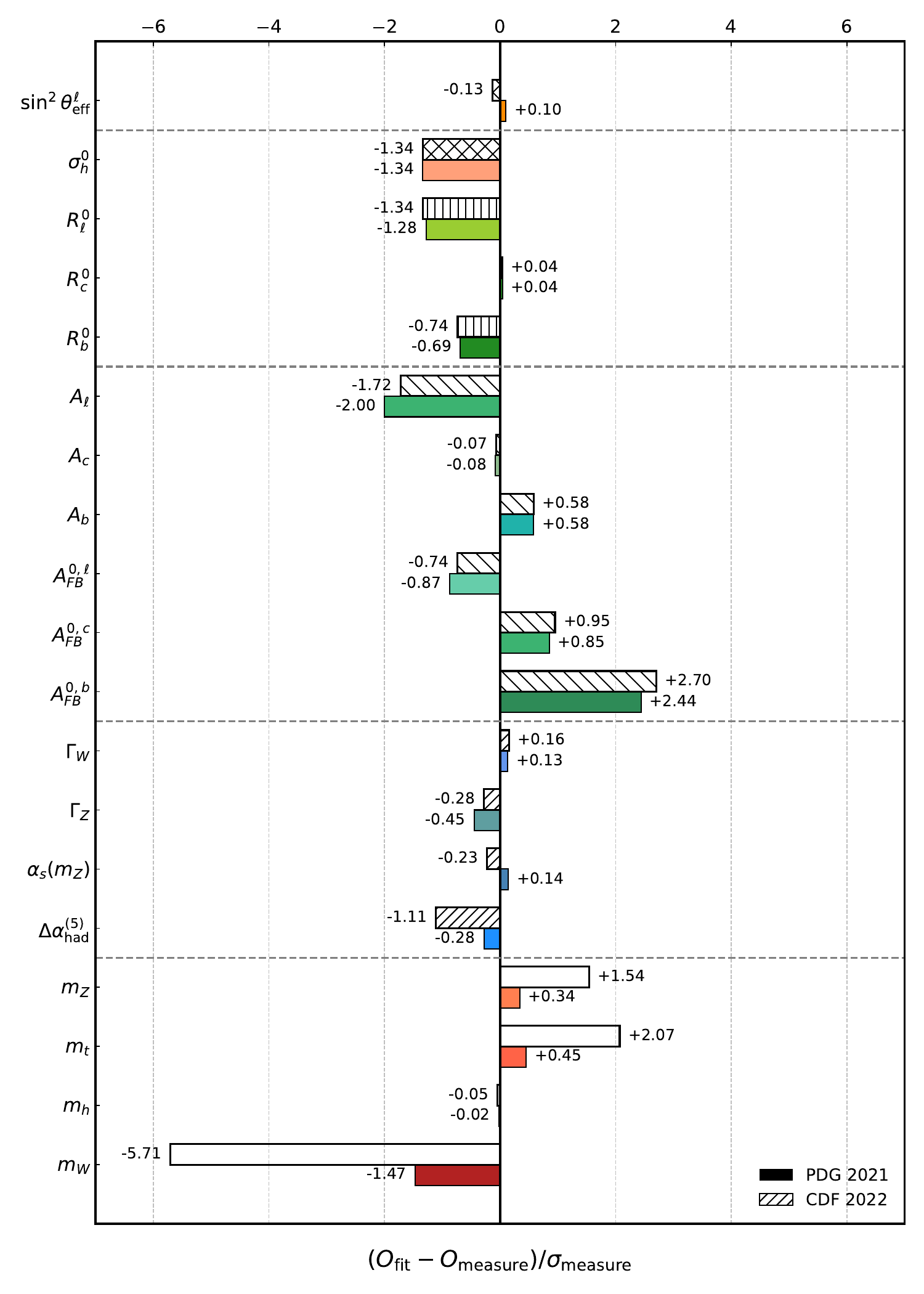}
	\caption{
		Comparison of pull values in global electroweak fits using the PDG 2021 value of the $W$ boson mass and the updated CDF 2022 measurement. The pulls are defined as $(O_{\mathrm{fit}} - O_{\mathrm{exp}})/\sigma_{\mathrm{exp}}$. The figure illustrates how the inclusion of the CDF result modifies the pull pattern across precision observables, inducing correlated shifts in $m_W$, $m_t$, $m_Z$, $A_{FB}^{0,b}$, and $\Delta\alpha_{\mathrm{had}}^{(5)}$. This behavior reflects the tension between the CDF measurement and the Standard Model electroweak fit, and supports the interpretation in terms of a global deformation of the fit requiring additional contributions, such as positive corrections to the oblique parameter $T$.
	}\label{fig:comparison_pdg_cdf_appendix}
\end{figure}
\clearpage
\begin{figure}[p]
	\centering
	\includegraphics[width=0.92\textwidth]{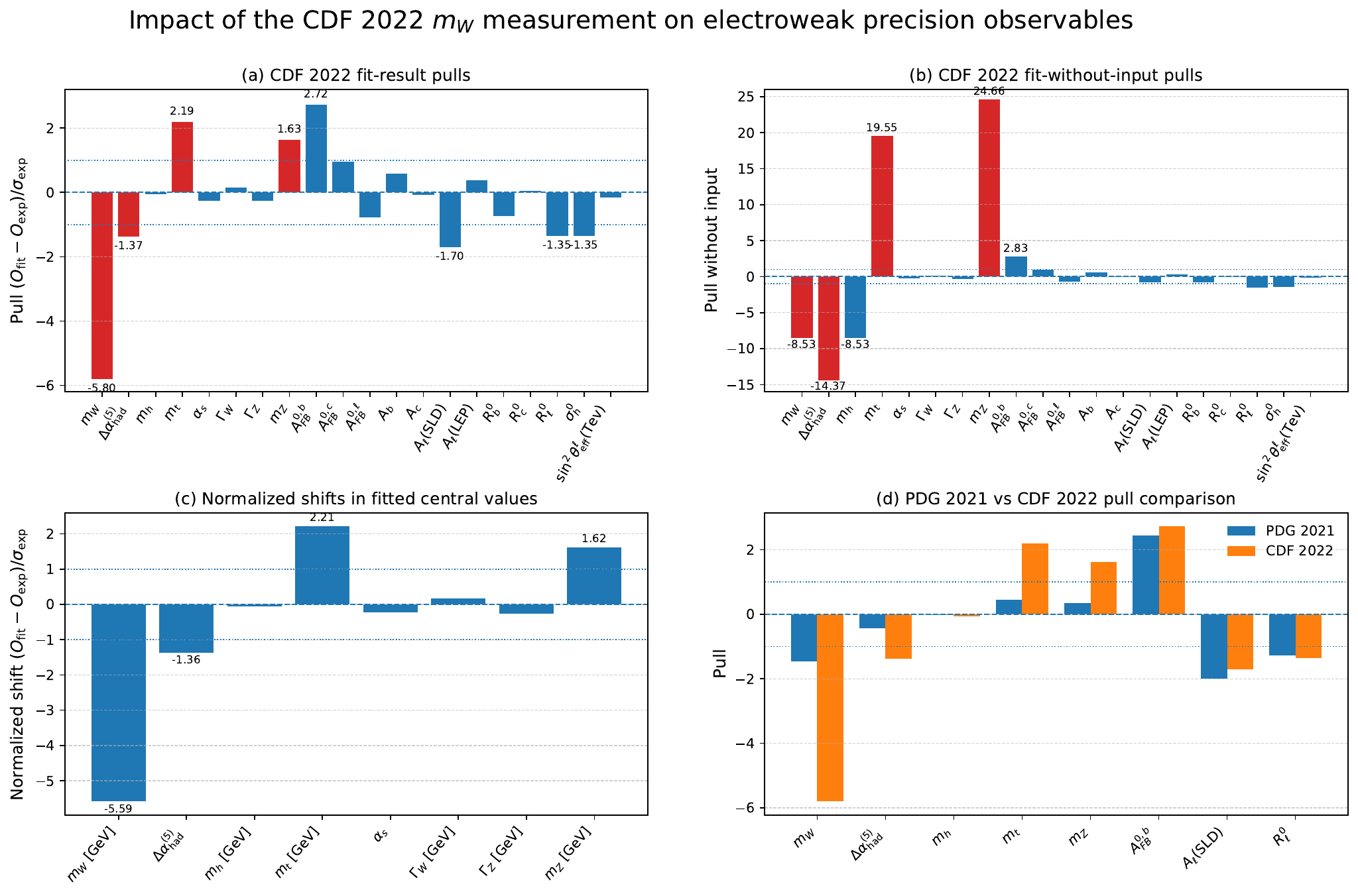}
\caption{
	Impact of the CDF 2022 $m_W$ measurement on electroweak precision observables, constructed consistently using the pull values displayed in Fig.~\ref{fig:EWfit}. Panel (a) shows the CDF 2022 fit-result pulls using the convention $(O_{\mathrm{fit}}-O_{\mathrm{exp}})/\sigma_{\mathrm{exp}}$, highlighting the dominant deviations in $m_W$, $m_t$, $A_{FB}^{0,b}$, $A_\ell(\mathrm{SLD})$, and $\Delta\alpha_{\mathrm{had}}^{(5)}$. Panel (b) displays the corresponding pulls obtained when each observable is removed from the fit input, thereby exposing the strong internal tension required to accommodate the CDF $m_W$ value within the Standard Model. Panel (c) presents the central-value shifts normalized to experimental uncertainties for a representative subset of observables, illustrating how the fitted parameters are displaced relative to their measured values. Panel (d) compares the PDG 2021 and CDF 2022 pull patterns, demonstrating how the inclusion of the CDF result induces a coherent deformation of the global electroweak fit. In particular, the large correlated shifts in $m_W$, $m_t$, $m_Z$, and $\Delta\alpha_{\mathrm{had}}^{(5)}$ indicate that the CDF measurement cannot be accommodated as an isolated effect, but instead points toward the need for additional positive oblique corrections, particularly in the $T$ parameter, which encodes custodial-symmetry breaking effects in extended scalar sectors.
}
	\label{fig:cdf_mw_impact}
\end{figure}
\clearpage
\twocolumngrid

\noindent\makebox[\textwidth][c]{\textbf{REFERENCES}}

\vspace{0.2cm}

\bibliographystyle{apsrev4-2}
\bibliography{refs}
\end{document}